\documentclass[%
 reprint, 
superscriptaddress,
nofootinbib,
 amsmath,amssymb,
 aps, physrev,
]{revtex4-2}
\usepackage[normalem]{ulem}
\usepackage{subcaption}
\usepackage{multirow}
\usepackage{tabularx}

\usepackage[ruled,vlined,linesnumbered]{algorithm2e}
\DontPrintSemicolon
\SetKwInput{KwInput}{Input}
\SetKwInput{KwOutput}{Output}
\usepackage{qcircuit}
\usepackage{braket}
\usepackage{amssymb}
\usepackage{hyperref} 
\usepackage{graphicx}
\usepackage{dcolumn}
\usepackage{bm}
\usepackage{float}
\usepackage{hyperref}
\usepackage[margin = 1in]{geometry}
\usepackage{lineno}
\usepackage{afterpage}
\usepackage{color}
\usepackage{amsmath}
\usepackage{nccmath} 
\usepackage{float}

\begin{document}

\title{Probing two-spin entanglement at quantum criticality on a quantum processor}

\author{Anshumitra Baul}
\email{baula@ornl.gov}
\affiliation{Quantum Science Center, Oak Ridge National Laboratory, Oak Ridge TN 37830, USA}
\affiliation{Quantum Information Science Section, Oak Ridge National Laboratory, Oak Ridge TN 37830, USA}

\author{Xiao Xiao}
\affiliation{Department of Physics, Northeastern University, Boston, Massachusetts 02115, USA}
\affiliation{Quantum Materials and Sensing Institute, Northeastern University, Burlington, Massachusetts 01803, USA}

\author{Phillip C. Lotshaw}
\email{lotshawpc@ornl.gov}
\affiliation{Quantum Science Center, Oak Ridge National Laboratory, Oak Ridge TN 37830, USA}
\affiliation{Quantum Information Science Section, Oak Ridge National Laboratory, Oak Ridge TN 37830, USA}

\date{\today}

\begin{abstract}

Quantum phase transitions in many-body systems give rise to highly entangled states, and understanding their quantum correlations is crucial for characterizing quantum materials. 
However, traditional entanglement measures such as entanglement entropy are difficult to interpret for noisy or mixed states and require complex circuits to evaluate. Therefore, we explore the Positive Partial Transpose (PPT) criterion, coupled with overlapping state tomography, as an efficient and scalable spin-spin entanglement witness. It detects pairwise entanglement from reduced density matrices, distinguishes quantum from classical correlations, and applies to both pure and mixed states. It is ideal for studying condensed matter systems prepared on noisy quantum devices as well as future extensions to finite temperatures. We demonstrate the approach on quantum hardware, using variational circuits to prepare quantum critical states with up to 20 qubits and completely map their two-spin entanglement across various quantum phase transitions.  
\end{abstract}
\maketitle
\begingroup
\renewcommand\thefootnote{} 
\footnotetext{This manuscript has been authored by UT-Battelle, LLC, under
Contract No. DE-AC0500OR22725 with the U.S. Department
of Energy. The United States Government retains and the publisher, by accepting the article for publication, acknowledges that
the United States Government retains a non-exclusive, paid-up,
irrevocable, world-wide license to publish or reproduce the published form of this manuscript, or allow others to do so, for the
United States Government purposes. The Department of Energy
will provide public access to these results of federally sponsored
research in accordance with the DOE Public Access Plan.}  
\endgroup

\section{Introduction}\label{sec1}

Entanglement has undergone a remarkable evolution since its original introduction in 1935 as a “spooky action at a distance”~\cite{einstein1935}.
Quantum entanglement lies at the heart of many exotic phenomena in quantum matter~\cite{Amico_2008,Horodecki1997}, from quantum spin liquids~\cite{Savary_2016}
and topological order~\cite{RevModPhys.89.041004}
to quantum criticality~\cite{Vidal2003} and thermalization~\cite{PhysRevLett.127.037201}. It is not only central to fundamental understanding in many-body physics but also essential for emerging quantum technologies ~\cite{horodecki2009,Nielsen2000}. The entangled quantum correlations surpass classical limits~\cite{brunner2014}, connecting the field of quantum information to the study of quantum phase transitions (QPTs). 

While classical phase transitions arise from thermal fluctuations, QPTs instead occur at zero temperature, where quantum fluctuations drive changes in the ground state as a control parameter is varied~\cite{Sachdev_2011,sondhi1997}, leading to nonanalytic changes in their properties including entanglement \cite{Osterloh_2002}. 
Despite its theoretical importance, characterizing entanglement in many-body systems remains a significant challenge~\cite{Amico_2008,Eisert2010}. Conventional classical methods—such as the Density Matrix Renormalization Group (DMRG)—face a fundamental bottleneck: their efficiency scales inversely with the amount of entanglement, causing computational resources to increase exponentially for states possessing the "volume law" entanglement found near critical points or in highly correlated phases. 

This directly motivates the transition to quantum hardware, which can simulate the highly entangled states that push classical simulations to their breaking point~\cite{Feynman1982,Nielsen2000,Adam_2024}. 
However, noise currently limits their operation, and there is an urgent need for standardized benchmarks to assess their efficiency and accuracy~\cite{boixo2018,Preskill2018}.  An essential aspect of this benchmarking is examining the extent of entanglement generated during the execution of quantum circuits~\cite{horodecki2009,Amico_2008}, providing a universal metric for hardware reliability\cite{Friis2019,Hamilton2022}. 

Entanglement witnesses  quantify entanglement in both critical states and quantum hardware. The most prominent example in condensed matter physics is the entanglement entropy (EE)~\cite{Amico_2008}.
EE is most directly interpreted for pure states, while noisy near-term quantum devices produce mixed states for which it is not a well-defined entanglement measure ~\cite{Cramer2010,Preskill2018}.
 
Existing techniques used in cold-atom and photonic platforms utilize twin-state interference to measure EE ~\cite{Islam2015,kaufman2016}. On digital quantum devices, however, this protocol effectively implements a controlled-SWAP operation between two identical copies of the state, which requires many controlled two- and three-qubit gates and is therefore highly noise-sensitive~\cite{foulds2021}. 
Instead, we prioritize local tomography to provide a more scalable framework for benchmarking subsystem entanglement in noisy, many-body systems. 

 Model-independent witnesses are also needed for large scale condensed matter  simulations. There has been recent interest in quantifying the basic entanglement properties of materials using alternative entanglement witnesses such as quantum Fisher information and measures based on few-spin correlations \cite{laurell2025witnessing,PhysRevLett.127.037201,hauke2016measuring}, which further motivates the approach we take here.

In this work, we implement a scalable framework for quantifying entanglement using the positive partial transpose (PPT) criterion \cite{Peres1996}. Unlike EE, PPT effectively detects pairwise qubit entanglement, in mixed, noisy and thermal states, making it suitable for near-term quantum devices. It can efficiently characterize all of the few-body entanglement in a state through efficient overlapping state tomography on quantum hardware \cite{Peruzzo2014}. We use the PPT to characterize correlations in 1D spin systems, offering a model-independent strategy to investigate highly entangled states and benchmark QPU performance~\cite{Peres1996,Horodecki1996}. We  employ matrix product states (MPS) to encode 1D spin chain ground states~\cite{verstrete2008,schollwockdmrg_2011} 
into variational quantum circuits
~\cite{Peruzzo2014,McClean2016} and enhance the accuracy of the hardware outputs through error mitigation. 
 This integration of PPT and variational quantum circuits  establishes a powerful workflow at the intersection of quantum information, condensed matter physics, and hardware validation.

\begin{figure*}[!htbp]  
    \centering  
    \includegraphics[width=1\linewidth]{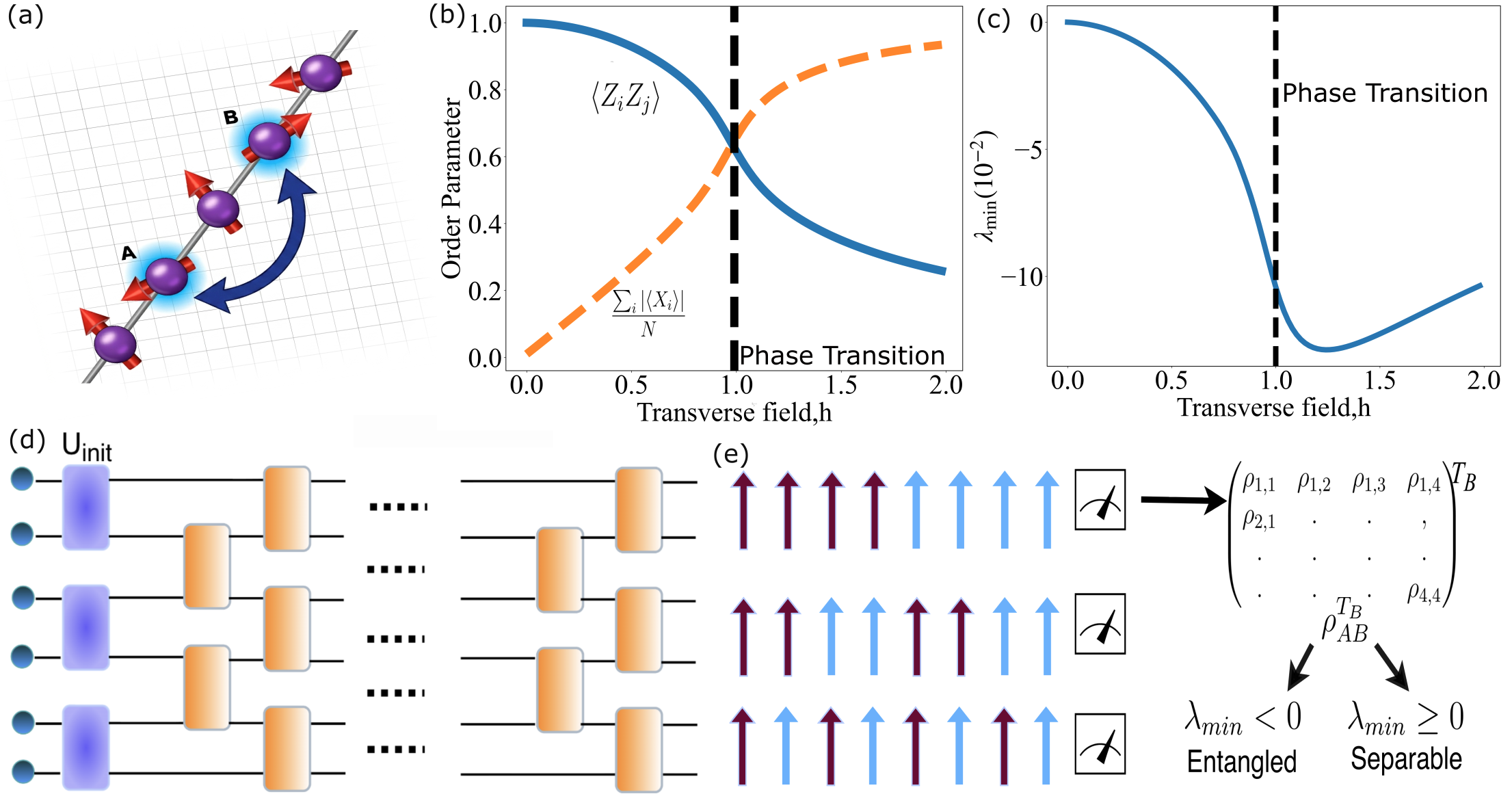}  
      \caption{\raggedright 
(a) Schematic 1D spin-model used to investigate two-spin entanglement on the quantum hardware.
(b) Evolution of the ferromagnetic $\langle Z_i Z_{i+1} \rangle$ and polarized phase $\sum_i|\langle X_i\rangle|/N$ order parameters across the TFIM QPT, with the critical value highlighted by a black dashed line. Results from MPS ground state calculations.
(c) PPT criterion of adjacent spins witnesses maximum entanglement near the critical point and less entanglement far from it.
(d) We prepare quantum critical states and evaluate their entanglement on quantum hardware using the variational brick-wall ansatz structure. 
(e) Schematic of the quantum overlapping tomography. Assigning different Pauli bases to the red(dark) and blue(light) spins we perform multiple joint product measurement settings given by the rows, to reconstruct $\rho_{AB}$ for all spin pairs $A$ and $B$. The resulting $\rho_{AB}$ are classified as entangled if the minimum eigenvalue of its partial transpose is negative ($\lambda_\text{min} < 0$) or separable if the eigenvalue is non-negative ($\lambda_\text{min} \ge 0$). 
}

\label{fig:loop_corr_ppt}

\end{figure*}
\section{Entanglement Witness and Quantum Critical Systems}\label{sec2}

\subsection{Entanglement}\label{sec2a}
Entanglement is a fundamental property of correlated quantum states and phases of matter that lack classical analogs~\cite{Sachdev_2011}. For a composite system $AB$ in a pure state $|\psi_{AB} \rangle$, the subsystems $A$ and $B$ are defined to be bipartite entangled if the state cannot be factored into independent local states, 
\begin{equation}|\psi_{AB} \rangle \neq |\psi_A \rangle \otimes |\psi_B \rangle.\end{equation} 
For $\ket{\psi_{AB}}$, the entanglement of $A$ with $B$ can be diagnosed through the reduced density matrix of either subsystem, which provides a complete characterization of the observable subsystem properties.  For $A$, this is defined through the partial trace over  $B$ as

\begin{equation} \rho_A = \mathrm{Tr}_B \ket{\psi_{AB}}\bra{\psi_{AB}} \end{equation} 
with an analogous expression for $\rho_B$.  If $A$ and $B$ are not entangled, then $\rho_A = \ket{\psi_A}\bra{\psi_A}$, while in the entangled case it takes the more general diagonal form
\begin{equation} \rho_A = \sum_i \lambda_i \ket{\psi_A^i}\bra{\psi_A^i} \end{equation}
The entanglement can be quantified using the entanglement entropy (EE)
\begin{equation} S_A = -\sum_i \lambda_i \ln \lambda_i, \end{equation}
with $S_A > 0$ signifying entanglement.  This is the standard account of the bipartite entanglement of pure states.

While EE is a well-defined 
measure for pure states, it is more difficult to interpret for mixed states, such as thermal states, states prepared on noisy quantum hardware, and individual spin pairs $A$ and $B$ within a larger chain~\cite{Amico_2008,hollands_2018,horodecki2009}. A broader framework is required to accurately characterize their quantum correlations. 
For example, the mixed state for the spin pair $AB$ shown in Fig.~\ref{fig:loop_corr_ppt}(a) is defined by the reduced density matrix, obtained by by tracing over all the other spins and expressed in diagonal form as
\begin{equation}\label{mixed rhoAB} \rho_{AB}= \sum_i \lambda_i |\psi^{i}_{AB} \rangle \langle \psi^{i}_{AB}|.\end{equation}
If $A$ and $B$ are in a pure state, we have only one nonzero diagonal eigenvalue, $\lambda_1 = 1$, while the state is mixed if $\lambda_i >0$ for multiple $\lambda_i$.

The definition of bipartite entanglement is more nuanced for mixed states such as (\ref{mixed rhoAB}) \cite{horodecki2009}. A mixed-state density matrix $\rho_{AB}$ is considered separable (not entangled) if it can be expressed as
\begin{equation} \label{separable rho}\rho_{AB} = \sum_{i} p_i \rho_A^{(i)} \otimes \rho_B^{(i)}.\end{equation}
Note that in this equation, the density matrix is not necessarily in diagonal form, and the components $\rho_A^{(i)}, \rho_A^{(i')}$, etc.~are not necessarily orthogonal. The condition (\ref{separable rho}) generalizes the idea of separability to include classical correlations in an ensemble. For example, the state $(\ket{0_A0_B}\bra{0_A0_B} + \ket{1_A1_B}\bra{1_A1_B})/2$ has classically-correlated states of $A$ and $B$, but  $A$ and $B$ are not entangled because the state is simply an ensemble of the separable components $\ket{0_A0_B}$ and $\ket{1_A1_B}$, neither of which possesses inherently-quantum correlations. If $\rho_{AB}$ cannot be written in the form (\ref{separable rho}),
\begin{equation} \label{entangled rho}\rho_{AB} \neq \sum_{i} p_i \rho_A^{(i)} \otimes \rho_B^{(i)},\end{equation}  
then subsystems $A$ and $B$ are entangled. We address entanglement across this mixed-state regime.

\subsection{Outline of the approach}\label{sec2b}

Fig.~\ref{fig:loop_corr_ppt} outlines the main ideas of this work. We implement a generic entanglement witness to quantify two-spin entanglement in quantum computations, taking quantum simulations of spin chains as a target example (Fig.~\ref{fig:loop_corr_ppt}(a)). The spin chains undergo a QPT as a control parameter, such as a magnetic field, is varied. The phase transition realizes a change in order parameter, for example as shown in Fig.~\ref{fig:loop_corr_ppt}(b). Near the transition point, the ground state becomes highly entangled. We argue the PPT criterion provides a practical and scalable witness of the resulting entanglement in quantum simulations, as seen in its peak near the phase transition point in Fig.~\ref{fig:loop_corr_ppt}(c). We demonstrate that quantum computers can scalably diagnose this entanglement by preparing these critical states on physical hardware using a structured circuit ansatz as shown in Fig.~\ref{fig:loop_corr_ppt}(d), to map the spin-chain topology onto the qubit register ~\cite{Baul2025}. Finally, Fig.~\ref{fig:loop_corr_ppt}(e) shows how we measure the reduced density matrices between every spin pairs, using quantum overlapping state tomography and quantify the entanglement generated at the hardware level for these many-body simulations. We explain each of these aspects in detail in the coming subsections.

\subsection{Positive Partial Transpose criterion}\label{sec2a}

 We detect entanglement by analyzing the partial transpose of the reduced density matrix $ \rho_{AB}$, which leads to the Peres-Horodecki criterion or the PPT test ~\cite{Peres1996}.

Given the reduced density matrix $\rho_{AB}$ in a composite Hilbert space $\mathcal{H}_A \otimes \mathcal{H}_B$, for two spins $A$ and $B$, represented as: \begin{equation} \rho_{AB} = \sum_{i,j,k,l} \rho_{ij,kl} |i\rangle_A \langle k| \otimes |j\rangle_B \langle l|, \end{equation} the partial transpose with respect to subsystem $B$ is defined as: \begin{equation} \rho_{AB}^{T_B} = \sum_{i,j,k,l} \rho_{ij,kl} |i\rangle_A \langle k| \otimes |l\rangle_B \langle j|. \end{equation} This operation swaps the indices of subsystem $B$ while keeping $A$ unchanged. The intuition behind performing this operation is that if the state is separable as in (\ref{separable rho}), then using Hermitian symmetry of $\rho_B^{(i)}$ the partial transpose will yield the complex-conjugated state $\rho^{T_B}_{AB} = \sum_i p_i \rho_A^{(i)}\otimes (\rho_B^{(i)} )^*$, which is still a valid quantum state. Alternatively, if the state is entangled as in (\ref{entangled rho}), then the partially-transposed density matrix may be unphysical.  The $AB$ entanglement can then be diagnosed through the eigenspectrum of $\rho_{AB}^{T_B}$.  If $\rho_{AB}^{T_B}$ has negative eigenvalues, then $\rho_{AB}$ could not have been separable, so it must be entangled.  

If all eigenvalues of $\rho_{AB}^{T_B}$ are non-negative, the state is separable (not entangled) in $2 \times 2$ and $2\times 3$ level systems\cite{Peres1996,Horodecki1996}. This is a necessary and sufficient condition for the two-spin entanglement. For a system of two spins $\rho_{AB}$, it is a proven property that the partially transposed density matrix $\rho_{AB}^{T_B}$ can have at most one negative eigenvalue $\lambda_\text{min}$, if entangled~\cite{verstraete2001}. The negativity $\mathcal{N}$ \cite{VidalWerner2002} is the absolute sum of the negative eigenvalues of $\rho^{T_B}_{AB}$. If there is a single negative eigenvalue $\lambda_{\min}<0$, this simplifies to
$$\mathcal{N}(\rho^{T_B}_{AB}) = |\lambda_{\min}|.$$
This criterion effectively detects entanglement in two or three level subsystems as described above.

We construct the two-qubit density matrix $\rho_{AB}$ from the hardware using basic ideas of quantum tomography. 
Using the Pauli basis, we can write the $\rho_{AB}$ as
\begin{equation}
    \rho_{AB} = \frac{1}{4} \sum
    \mathrm{tr}\!\left[(\sigma_A^\alpha \otimes \sigma_B^\beta)\,\rho\right] 
    \sigma_A^\alpha \otimes \sigma_B^\beta,
\end{equation}
(where $\hat{\sigma}^{\alpha}(\alpha=\mathbb{I},x,y,z)$ are the Pauli matrices) expanding in terms of all the correlation functions. We applied the quantum overlapping tomography (QOT) technique described in Ref.~\cite{Cotler_QOT_2020} to effectively calculate all the two-qubit reduced density matrices. As illustrated in Fig.~\ref{fig:loop_corr_ppt}(e), this process begins by performing local measurements on various spin configurations within the chain to reconstruct the reduced density matrix $\rho_{AB}$ for specific pairs. QOT requires a total of $M=O(\log N)$ measurement settings to determine the two-point correlations for all spin pairs efficiently.
The method can also be extended to systematically map $k$-body correlations using $\mathcal{O}(3^k \log^2(N))$ measurements \cite{Cotler_QOT_2020}, though we will not pursue these higher-order correlations here. Alternative approaches to efficiently characterizing few-body correlations are presented in Refs.~\cite{huang2020predicting,mukherjee2025}.

To diagnose entanglement, for each spin pair $AB$ we perform a partial transpose on the second subsystem to obtain $\rho_{AB}^{T_B}$. And following the PPT criterion, we confirm the presence of entanglement if $\lambda_\text{min} < 0$, whereas $\lambda_\text{min} \geq 0$ indicates a separable state.  

We seek statistically significant entanglement signatures, requiring uncertainty quantification for $\lambda_\text{min}$. Since $\lambda_\text{min}$ is obtained by reconstructing a two-spin reduced density matrix and diagonalizing its partial transpose, it depends nonlinearly on the measured correlators, with uncertainty that is difficult to quantify through error propagation. Therefore, we use bootstrap resampling to estimate the uncertainty of $\lambda_\text{min}$. We use $1000$ bootstrap samples of the QOT measurement results, where in each of these samples we compute each correlator using the same number of shots as in our raw results, but here with the shots drawn from the set of measurement results with repeated draws allowed. The standard error of $\lambda_\text{min}$ is taken as the standard deviation of the bootstrap means, and we consider $\lambda_\text{min}$ that are more than one standard error below zero to be statistically-significant entanglement witnesses. 

\subsection{Spin Hamiltonians}\label{sec2b}

 This section details the one-dimensional Hamiltonians used to demonstrate the PPT criterion, in all cases using periodic boundary conditions. The Transverse-Field Ising Model (TFIM) and the XXZ model are canonical systems for characterizing QPTs across their respective control parameters. The choice of an appropriate order parameter to characterize a phase transition is often not straightforward, except in a few well-understood cases \cite{PFEUTY197079}. As QPTs are driven by ground-state fluctuations, we examine local observables and spin–spin correlations across the lattice to capture how these fluctuations reorganize near criticality. Previous literature has successfully identified regions of high correlation in these models using EE~\cite{Osborne2002, Vidal2003, ar2008}. We complement these established results by characterizing the mixed-state pairwise entanglement within these critical regions, providing a more granular view of the correlation structure.

\subsubsection{Transverse Field Ising Model(TFIM)}\label{sec2b1}
The Ising model was designed to determine whether or not local interactions between magnetic spins could produce a macroscopic net magnetic moment \cite{Ising1925}. We consider an Ising model with transverse field \cite{PFEUTY197079, Stinchcombe_1973}.  
The Hamiltonian is given as
\begin{equation}
    \hat{\mathrm{H}}_{\text{TFIM}}=-J\sum^{N-1}_{i=1}\sigma^{z}_{i}\sigma^{z}_{i+1}+h\sum^{N}_{i=1}\sigma^{x}_{i},
\end{equation} 
where $J=1$ is the coupling between the spins and $h$ is the transverse field applied in the $x$-direction. When $J>0$, the system is ferromagnetic, and antiferromagnetic when $J \leq 0$.  At the critical point, it undergoes a second-order quantum phase transition as shown in Fig.~\ref{fig:loop_corr_ppt} (b).  
The local correlation function $\langle Z_i Z_{i+1} \rangle$ and average magnetization $\sum_{i} |\langle X_{i}\rangle|/N$, shown in Fig.~\ref{fig:loop_corr_ppt}(b), serve as signatures of QPT between the ferromagnetic and polarized phases.

We set $J=1$ and prepare states across the transverse field range $h \in [0.0, 2.0]$.  
\subsubsection{XXZ model}\label{sec2b2}

The spin-$1/2$ XXZ model is defined by the Hamiltonian 
\begin{equation}
    \hat{\mathrm{H}}_\text{XXZ}=-J\sum^{N}_{i=1}( \sigma^x_{i} \sigma^{x}_{i+1}+\sigma^y_{i} \sigma^{y}_{i+1}) + \Delta \sum^{N}_{i=1}\sigma^z_{i} \sigma^{z}_{i+1}
\end{equation}
$\Delta$ is the anisotropy parameter. 
The system is in a gapless phase for $-1 < \Delta \leq 1$, exhibiting highly entangled states. It undergoes a first-order phase transition to the ferromagnetic phase at $\Delta =-1$, while at $\Delta =1$ there is an inifinite-order Kosterlitz-Thouless Quantum phase Transition to the antiferromagnetic phase. 

The XXZ model cannot be diagonalized analytically, but its energy spectrum can be obtained by Bethe ansatz ~\cite{PhysRev.150.321}. We set for $J=1$ and prepare ground states across the anisotropy range $\Delta \in [-1.5,1.5]$.

\subsubsection{Quantum state preparation }\label{2b3}

To prepare quantum states at and near the expected critical points, we utilize a family of physics-motivated variational ans\"atze featuring a state-initialization layer ($\mathrm{U}_{\mathrm{init}}$) followed by depth-$L$ layers of one and two-qubit gates arranged in a brickwork pattern, as shown in Fig.~\ref{fig:loop_corr_ppt}(d)~\cite{mukherjee2026}. This circuit topology is designed to capture the local and global correlation structures of 1D quantum spin chains while maintaining a gate depth suitable for noisy hardware~\cite{yu2023}. 
Additional details are in appendix~\ref{appendix1}, we present a summary of the approach below. 

The ansatz uses a structure based on adiabatic evolution from valence bond ground state (prepared by $\mathrm{U}_{\mathrm{init}}$) and evolving toward the target ground state of the system. 
 The idealized noiseless state prepared by our variational ansatz is 
\begin{equation}
|\psi_{\mathrm{ansatz}}\rangle = \mathrm{U}(\boldsymbol{\theta}) |0\cdots 0\rangle = \mathrm{U}_{\mathrm{var}}(\boldsymbol{\theta})\, \mathrm{U}_{\mathrm{init}} |0\cdots 0\rangle,
\label{eq:state}\end{equation}
where $\mathrm{U}_{\mathrm{var}}(\boldsymbol{\theta})=\Pi^{L}_{n=1}[\mathrm{U}^{(n)}_{\mathrm{even}}({\theta_{e}})\mathrm{U}^{(n)}_{\mathrm{odd}}({\theta_{o}})] $, where $\mathrm{U}^{(n)}_{\mathrm{even}}({\theta_{e}})$ and $\mathrm{U}^{(n)}_{\mathrm{odd}}({\theta_{o}})$ are related to the interactions of the Hamiltonian. This approach is well-suited for variational optimization because, in principle, the ansatz can adiabatically prepare the ground state exactly if a sufficient number of layers $L$ are employed. We focus on the total qubit number $N$ being even, because a product of valence bond state can be adiabatically connected to the ground state of the model Hamiltonian for finite even $N$.

\subsubsection{Analysis of simulated circuit optimization}\label{sec2b3}

We optimize our ground state preparation circuits using qubit sizes $N=12,20$ that are consistent with periodic linear qubit arrays on IBM hardware.

We maximize the fidelity $|\langle \psi_\text{ansatz} (\bm{\theta})| \psi_\text{gs} \rangle|^2$ with respect to the true ground state $\ket{\psi_\text{gs}}$, with energy $E_\text{gs}$, computed from DMRG, to obtain the optimal parameters $ \bm{\theta}^{*} $. We then compute the ansatz energy, $E_\text{ansatz}( \bm{\theta} ^{*} )= \langle \psi_\text{ansatz}( \bm{\theta^{*}} )| \hat{\mathrm{H}} |\psi_\text{ansatz}( \bm{\theta^{*}} ) \rangle $ and its  error $\epsilon = |E_\text{ansatz}( \bm{\theta ^{*}} )- E_\text{gs}|/|E_\text{gs}|$.  
 
We present the numerical details for the state preparation circuits in Table~\ref{tab:circuit}. 

\begin{table}[t]
\centering
\small
\setlength{\tabcolsep}{11.4pt}
\begin{tabularx}{\columnwidth}{|c|c|c|c|c|}
\hline
Model & $N$ & $L$ & Fidelity & $(1-\epsilon)\%$ \\
\hline
\multirow{2}{*}{TFIM} & 12 & 4  & 0.998 & 99.83 \\
\cline{2-5}
                       & 20 & 6  & 0.992 & 99.8  \\
\hline
\multirow{2}{*}{XXZ}  & 12 & 6  & 0.989 & 99.21 \\
\cline{2-5}
                       & 20 & 10 & 0.982 & 98.15 \\
\hline
\end{tabularx}

\caption{Ground state preparation circuit results for the TFIM and XXZ models. Refer to Fig.~\ref{fig:layer_fid_energy} for more details.}
\label{tab:circuit}
\end{table}

The TFIM is integrable and its critical state is relatively easy to prepare, while the XXZ model involves $U(1)$ symmetry and more intricate exchange interactions. The increased density of entanglement and the nature of the correlations in the XXZ model typically require a deeper circuit to capture the ground state, even for a smaller number of sites. It is hard to prepare states at the quantum phase transition due to long range quantum correlations, seen in Fig.~\ref{fig:layer_fid_energy}(a, c). The number of layers significantly increase near the phase transitions for both the models.

\section{Results}\label{sec3}
We performed experiments by preparing the TFIM and XXZ ground states on the
156-qubit IBM Quantum \texttt{ibm\_boston} backend, accessed through the cloud,
using $20{,}000$ shots for each circuit. The \texttt{ibm\_boston} device is a
Heron r3 superconducting quantum processor with heavy-hexagonal connectivity.
The hardware-executed circuits were the optimized circuits obtained in section\ref{sec2b3}. These circuits were then passed through Qiskit's preset pass
manager to make them compatible with the backend topology and native gate set,
including layout selection, routing, and basis-gate translation. We did not use
Qiskit's circuit-optimization passes to further reduce or modify the optimized
ansatz circuits.

\subsection{Probing entanglement on quantum hardware}\label{sec3a}

To evaluate hardware performance under realistic conditions, we utilize quantum spin models—the TFIM and XXZ chains—as benchmark systems for exploring entanglement at phase transitions \cite{Stinchcombe_1973}. We evaluate entanglement using  the PPT criterion directly from raw measurement counts obtained from the 156-qubit IBMQ Boston, assessing the device's capacity to generate and sustain entanglement in the presence of noise \cite{Friis2019, Hamilton2022}. 

\begin{figure*}[!htbp]
\centering

\begin{subfigure}{0.48\textwidth}
\centering
\includegraphics[width=\linewidth]{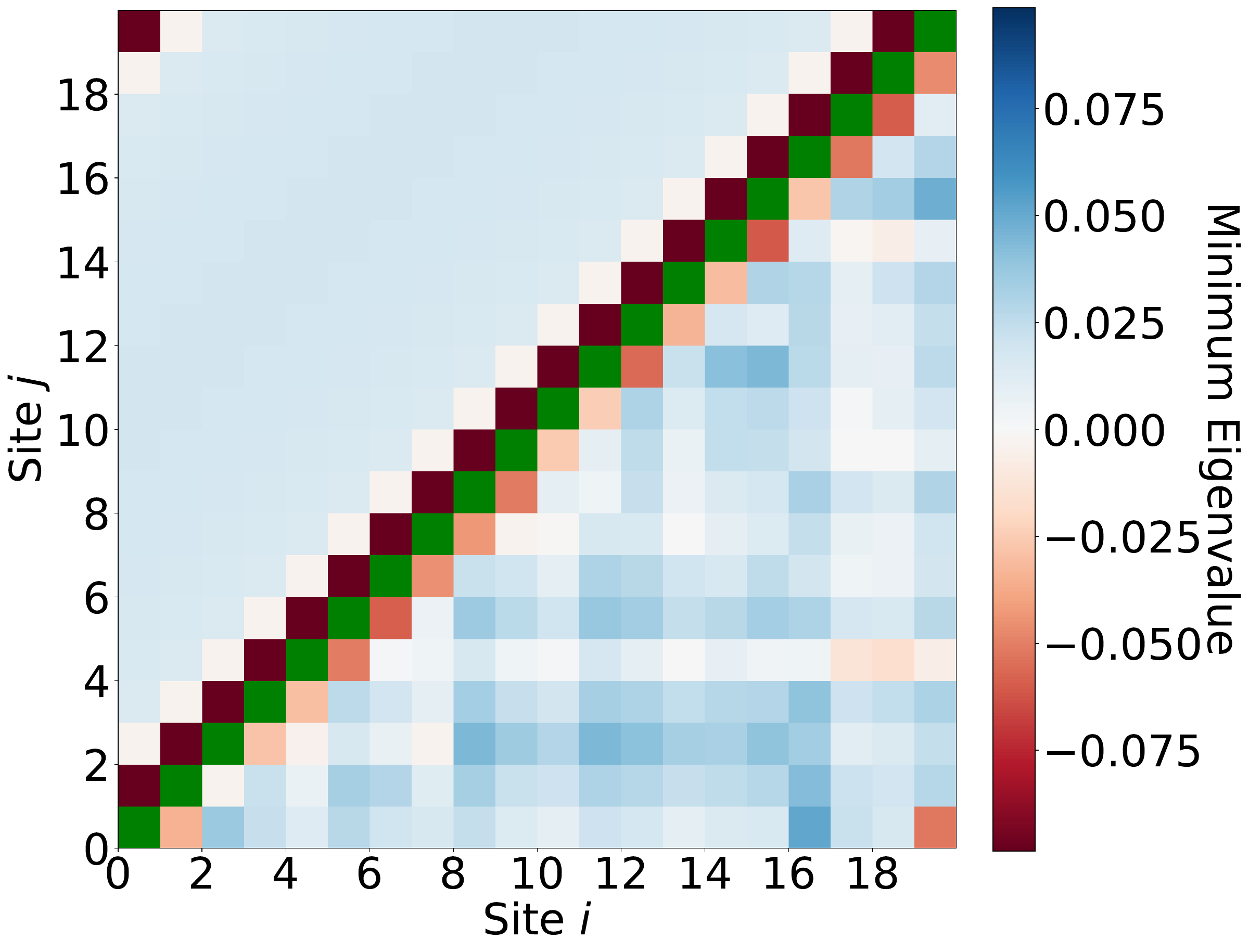}
\caption{}
\end{subfigure}
\hfill
\begin{subfigure}{0.47\textwidth}
\centering
\includegraphics[width=\linewidth]{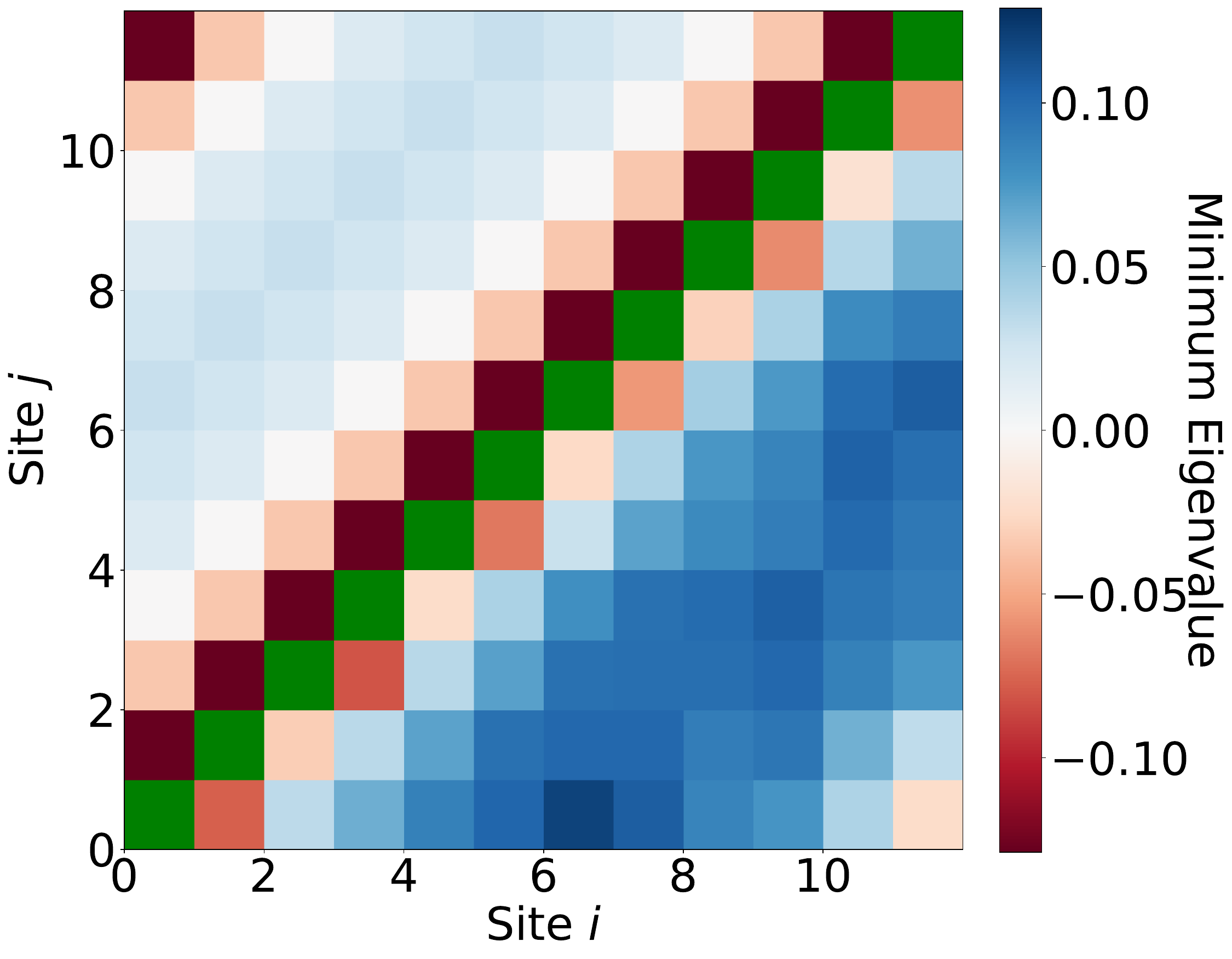}
\caption{}
\end{subfigure}
\vspace{0.3cm}
\begin{subfigure}{0.48\textwidth}
\centering
\includegraphics[width=\linewidth]{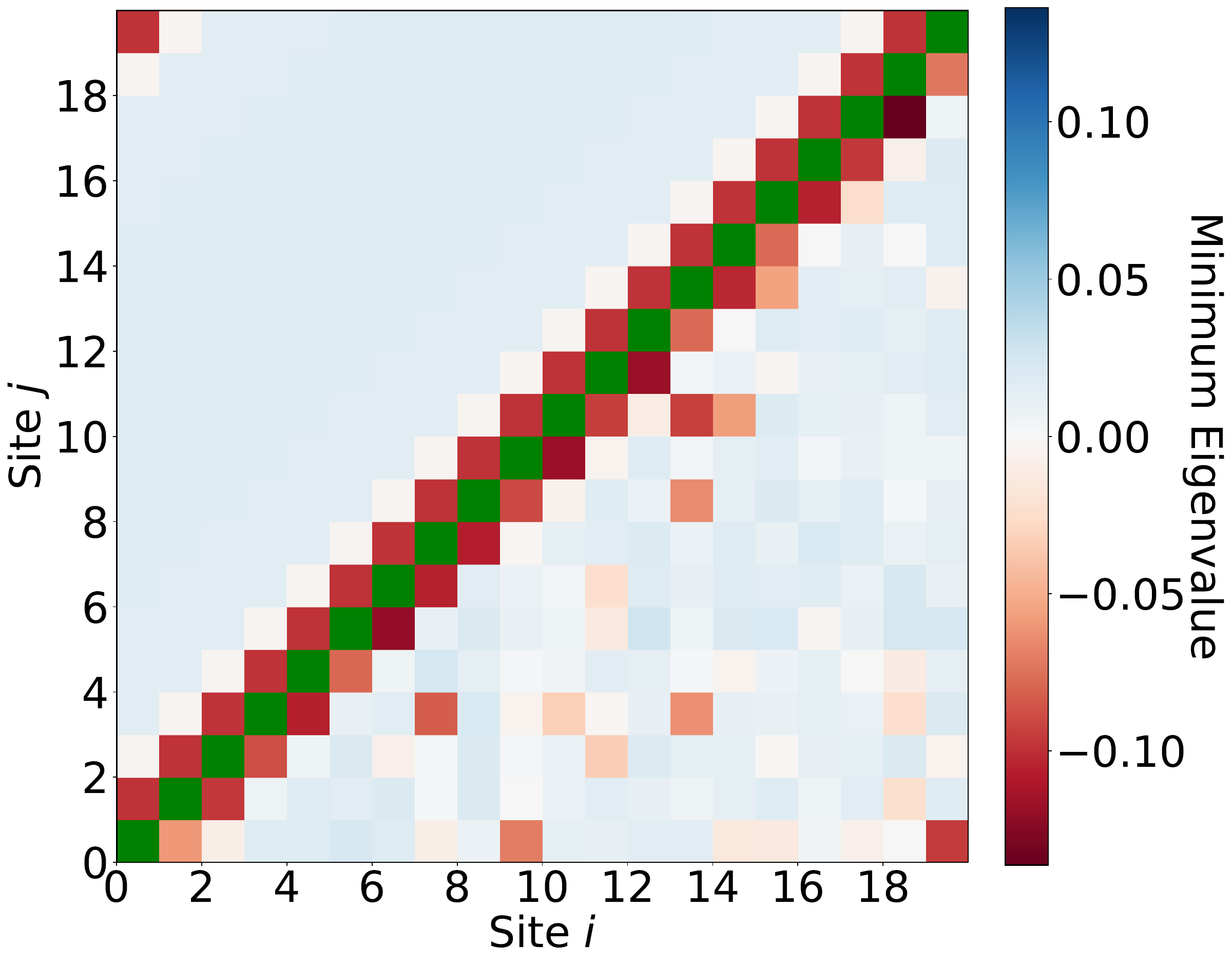}
\caption{}
\end{subfigure}
\hfill
\begin{subfigure}{0.47\textwidth}
\centering
\includegraphics[width=\linewidth]{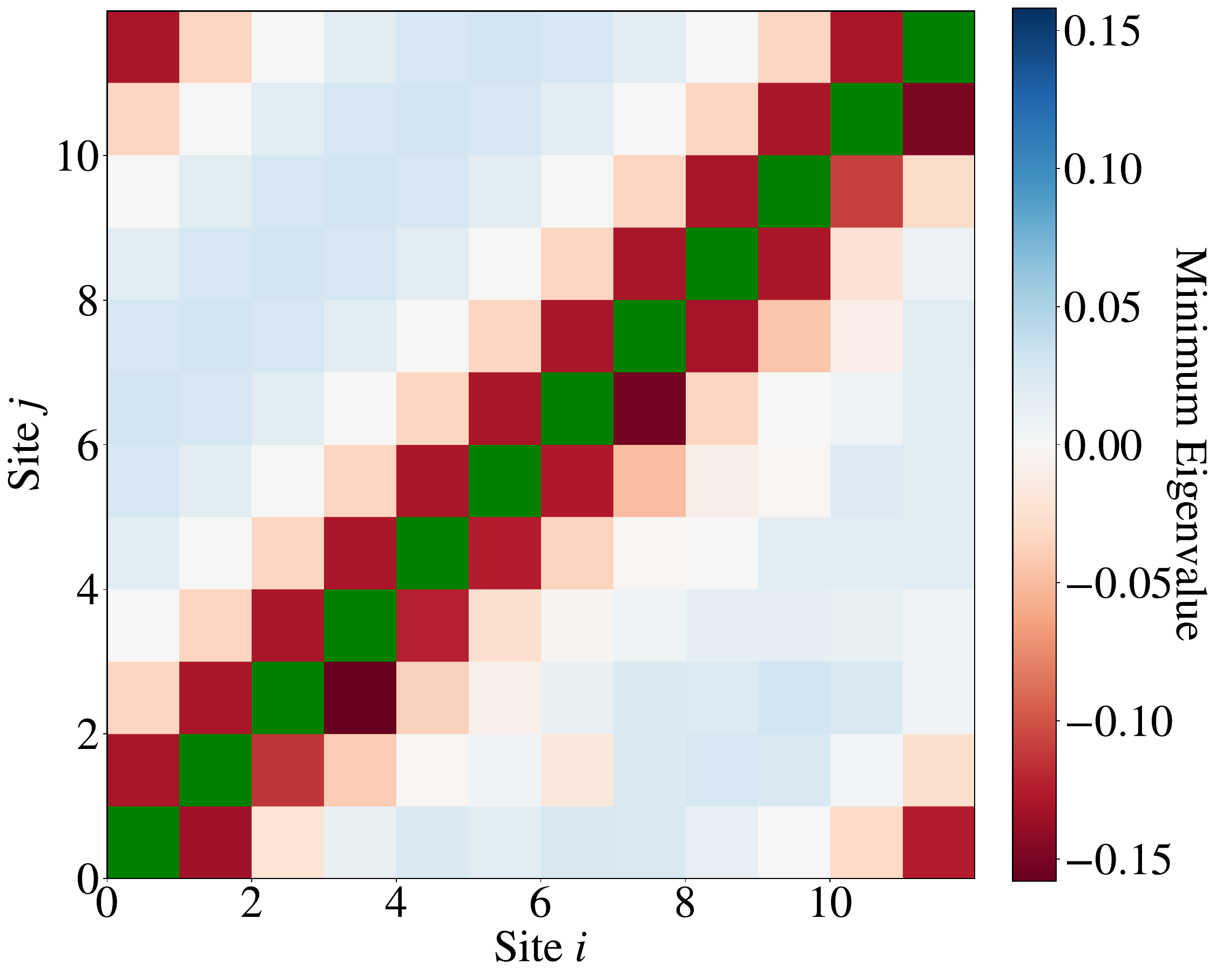}
\caption{}
\end{subfigure}

\caption{\raggedright Heatmaps of the PPT entanglement witness for all spin pairs $(i, j)$ near the critical points of the 1D systems, with $\lambda_\text{min} < 0$ (orange,red) indicating entanglement. The upper panels show the unmitigated $\lambda_\text{min}$ and the lower panels show the $\lambda_\text{min}$ after error mitigation. (a,c) TFIM at $h = 1.0$ and (b,d) XXZ at $\Delta = -0.68$. The plots are partitioned into two regions for direct benchmarking: the upper-left triangle  ($j > i$) represents the exact results obtained from MPS calculations, while the lower-right triangle ($i > j$) displays experimental results measured on quantum hardware. The green diagonal corresponds to identical sites ($i=j$), where a pairwise reduced density matrix cannot be defined. The hardware captures the strong nearest-neighbor correlations immediately adjacent to the diagonal. The unmitigated data fails to resolve the weaker long-range entanglement between next-nearest neighbors in the XXZ model; however, after applying error-mitigation techniques, the mitigated results agree well with the MPS benchmarks. Note that all spurious long range entanglement in the TFIM hardware results is not statistically significant.
}
\label{fig:ppt_raw}

\end{figure*}

Fig.~\ref{fig:ppt_raw} compare the spatial structure of the hardware-derived minimum eigenvalues against ideal MPS benchmarks, where negative values (red/orange) identify entangled spin pairs. The upper-left triangle of each plot displays the ideal MPS benchmarks, while the lower-right triangle shows the unmitigated hardware measurements in Fig.~\ref{fig:ppt_raw}(a),(b), allowing for a direct site-by-site comparison of the entanglement structure. We observe that the hardware can resolve the strong nearest-neighbor entanglement next to the diagonal. But it generally overestimates the eigenvalues due to noise, resulting in a "bluer"  plot where next-nearest-neighbor entanglement present in the MPS results is prematurely lost, particularly in the XXZ model. In the hardware results, the spurious long-range entanglement in the TFIM model is not statistically significant. 

We have seen that extracting $\lambda_\text{min}$ from shot-based data provides a practical way to quantify the entanglement of many-body states on quantum processing units, but to better align our results with exact theoretical values we need to mitigate the hardware noise.

\subsection{Error Mitigation}\label{sec3b}

Here we describe the approaches we have used to implement error mitigation on readout and gate errors.

\subsubsection{Measurement (Readout) Error Mitigation}\label{sec3b1}
Readout error 
 
mitigation is useful for recovering accurate probability distributions from measured counts. 

To characterize how the measurement process maps ideal input states to observed outcomes, 
we estimate the measurement assignment matrix $\mathcal{M}$ satisfying
\begin{equation}
    \vec{P}_{\mathrm{measured}} = \mathcal{M} \, \vec{P}_{\mathrm{ideal}},
\end{equation}
where $\vec{P}_{\mathrm{measured}}$ and $\vec{P}_{\mathrm{ideal}}$ denote the measured and ideal probability distributions, respectively. 
We obtain a mitigated distribution suitable for evaluating observables, by inverting this relation. For an $N$-qubit system, the full matrix $\mathcal{M}$ has dimension $2^N \times 2^N$; 
this scaling makes full readout calibration impractical for large systems. Instead, we use the matrix-free measurement mitigation (M3) routine that avoids the usual
exponential overhead by  
working in a subspace of the noisy input bit strings, as described in Ref.~\cite{M3_neereja}.  This approach focuses on the subset of the bitstrings with the largest observed probabilities actually measured in the experiment, allowing it to function effectively in a reduced subspace rather than the full $2^N$ Hilbert space. 

Readout mitigation allows us to better characterize the state prepared on the quantum device. However, the expectation values of observables are still influenced by gate errors, which we address in the next subsection.

\subsubsection{Zero-noise Extrapolation (ZNE)}\label{sec3b2}

Several earlier works have addressed the gate errors by extrapolating measured observables to the zero-error limit ~\cite{Endo2018,Temme2017,Kandala2019}. This method corrects the expectation values themselves rather than the underlying quantum state.

 We start with the variational ansatz U given in Eq.~\ref{eq:state}. 

The approach builds on the technique introduced in Ref.~\cite{GiurgicaTiron2020}, where they construct a family of noise-amplified circuits by
inserting additional digital quantum gates that prepare the states
\begin{equation} \label{ZNE basic}
    |\psi_{n}\rangle = \bigl[\, \mathrm{U}(\mathrm{U^\dagger}\mathrm{U})^{\,n} \bigr] |0\cdots 0\rangle,
\end{equation}
where $n$ is a nonnegative integer, referred to as the number of folds of $\mathrm{U}$. Then the target observable is measured
\begin{equation} \label{On}
    \mathcal{O}_{n} = \langle \psi_{n} | \hat{\mathcal{O}} | \psi_{n} \rangle
\end{equation}
for several values of $n$. 
In an ideal, noise-free device, all $\mathcal{O}_n$ would be identical because inserting $\mathrm{U}^{\dagger}\mathrm{U}$ does not change the state. In practice, increasing $n$ amplifies the effect of gate noise, degrading $|\psi_n\rangle$ and $\mathcal{O}_n$. 
To estimate the zero-noise value, the quantity $\mathcal{O}_{n}$ is fit as a function of the effective noise-scaling factor $\lambda = 2n + 1$ relating to the number of occurrences of U in (\ref{On}) (not to be confused by minimum eigenvalue, $\lambda_\text{min}$) and extrapolated to the limit $\lambda \rightarrow 0$.

The approximate influence of noise can be understood using a very simple error model in which global depolarizing channels are applied with each gate. In this model,  
the error rate compounds after $n$ operations, with a depolarizing error parameter $p$ that scales exponentially with the number of repetitions~\cite{Nielsen2000}. 
 The noisy state $\rho_n(\lambda)$ after $n$ operations is expressed as a mixture of the ideal state and the maximally mixed state $\mathbf{I}/2^N$, 
\begin{equation}\label{depolarized rho} \rho(\lambda) = p^\lambda \rho_{ideal} + (1 - p^\lambda) \frac{\mathbf{I}}{2^N},\end{equation}
with $\mathbf{I}$ the identity operator. The expectation value of an observable $M$, given by  $M(\lambda) = \text{Tr}(M \rho(\lambda))$, inherits this exponential dependence. By defining $p^\lambda = e^{\lambda \ln p}$, we map the physical noise scaling to a standard exponential decay ansatz, 
\begin{equation}\label{fM} f_{M}(\lambda) = a e^{-b \lambda} + c,\end{equation}
where $c$ represents the expectation value in the limit of infinite noise (the maximally mixed state), and $a+c$ corresponds to the noise-free expectation value at $\lambda = 0$. For our fitting, we assign $c=0$ in Eq. (\ref{fM}) so $f_{M}(\lambda) \rightarrow 0$ as $\lambda \rightarrow \infty$, as expected for Pauli operator expectations in our simple noise model (\ref{depolarized rho}). The approach works well for both the TFIM and XXZ models on the hardware. 

There are a few variations of how folding is applied; global folding as in Eq. (\ref{ZNE basic}) amplifies noise by folding the full circuit as $\mathrm{U}\,\mathrm{U}^\dagger\,\mathrm{U}$, while local folding  applies the inverse operations $\text{U}_\text{g}^\dag \text{U}_\text{g}$ to each gate $\text{U}_\text{g}$ in $\mathrm{U} $ separately. These variations can produce different noisy signals in real hardware, where coherent errors, crosstalk, and time-correlated/non-Markovian effects~\cite{Temme2017,vega2017,sarovar2020} are important. 

While standard global or local folding produces only odd integer noise scale factors, these large factors can significantly increase the noise in the measured signal, yielding poor results for extrapolation from our circuits. To improve extrapolation accuracy, we use a technique called partial folding to incorporate even and fractional noise scale factors, mentioned in Ref.~\cite{majumdar2023}.

With partial folding, the original gates are not all folded
an equal number of times. 
Instead, we first extract all gate operations from the quantum circuit and store them in an ordered list. Using this list, we construct the full unitary operator $\mathrm{U}$ and its inverse $\mathrm{U}^\dagger$. 
We then divide $\mathrm{U}$ into smaller chunks and construct a partial unitary $\mathrm{U}_{\text{par}}$ and its adjoint $\mathrm{U}_{\text{par}}^\dagger$ from selected chunks. These partial folds are inserted to reach the desired scale factor. \par
For example, for scale factor $1.5$, we divide the $\mathrm{U}$ into four chunks as $\mathrm{U} = \mathrm{U}_1 \mathrm{U}_2 \mathrm{U}_3 \mathrm{U}_4$, build $\mathrm{U}_{\text{par}}=\mathrm{U}_{i}$ and $\mathrm{U}^{\dagger}_{\text{par}}=\mathrm{U}^{\dagger}_{i}$, where $i$ is from one of the four chunks $\mathrm{U}_i$. If $i=1$ we apply $\mathrm{U}_{1}\mathrm{U}^{\dagger}_1 \mathrm{U}_1 \mathrm{U}_2 \mathrm{U}_3 \mathrm{U}_4$ and similarly for the other choices of $\mathrm{U}_\text{par}$. We have four sample circuits for scale factor $1.5$, and average over those. \par
We use a similar approach for all $\lambda \in \{1.5,2.0,2.5,3.0,3.5,4.0,4.5,5.0\}$, always partitioning $\mathrm{U}$ into four chunks $\mathrm{U}=\mathrm{U}_1 \mathrm{U}_2 \mathrm{U}_3 \mathrm{U}_4$ and constructing the noise-scaled circuit by inserting $\mathrm{U}_i^\dagger \mathrm{U}_i$ folds at chunk boundaries. For each $\lambda$ we generate a circuit ensemble by choosing which chunks are partially folded with a fixed number of shots, and then average over the resulting set of circuits. For example, for $\lambda=2$ there are six fold variants that we average over.

We find this gives a much more satisfactory fit than the simple global and local folding approaches, as we describe in Sec.~\ref{sec3c}. 

\subsubsection{Symmetry of the Hamiltonians}\label{3b3}

Symmetry constraints within the two Hamiltonians directly eliminate several correlation functions to reconstruct the reduced density matrices $\rho_{AB}$. In the TFIM, the global parity operator $\prod_i X_i$ is a symmetry of the Hamiltonian and its eigenstates, which ensures that terms like $\langle XY \rangle$, $\langle XZ \rangle$, $\langle ZI \rangle$, $\langle YI \rangle$ and $\langle YX \rangle$ vanish \cite{Sachdev_2011}. We therefore focus exclusively on the non-vanishing $\langle ZZ \rangle$, $\langle XX \rangle$, $\langle YY \rangle$ and $\langle XI \rangle$ components, setting the others equal to zero.

Similar to the TFIM, the XXZ model exhibits symmetries that restrict the set of non-vanishing observables. Beyond the discrete $Z_2$ symmetry, the XXZ Hamiltonian possesses a continuous $U(1)$ symmetry representing rotational invariance around the $z$-axis. The expectations $\langle XY \rangle$, $\langle XZ \rangle$, $\langle ZI \rangle$, $\langle XI \rangle$, $\langle YI\rangle$, and $\langle YX \rangle$ vanish identically, so we calculate only $\langle XX \rangle $, $\langle YY \rangle$, and $\langle ZZ \rangle $ correlations for our error-mitigated results.

\subsection{Cloud Experiment Results}\label{sec3c}

\subsubsection{Transverse Field Ising Model(TFIM)}\label{sec3c1}
\begin{figure*}[!htbp]  
    \centering  
    \includegraphics[width=1\linewidth]{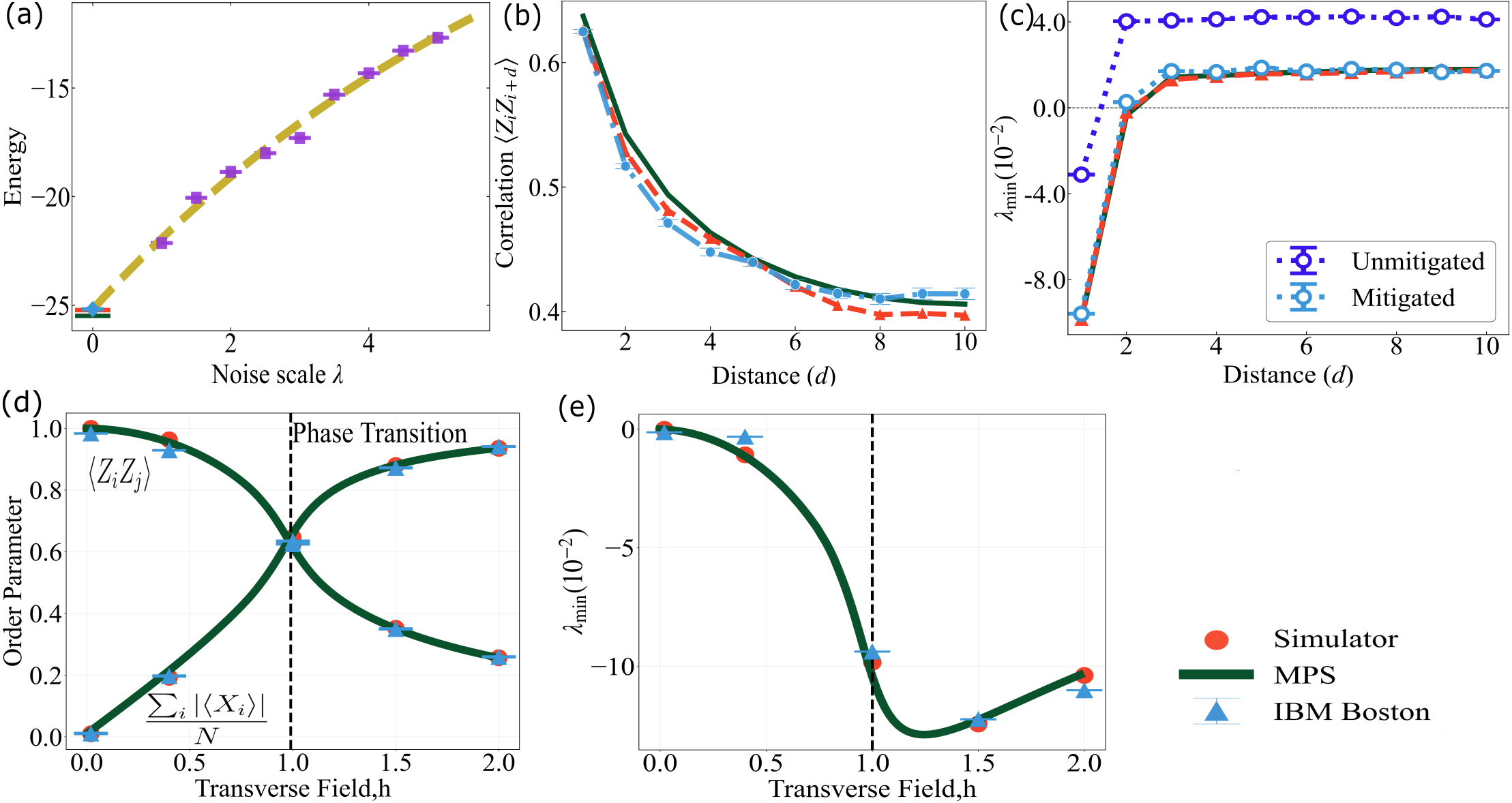}  

      \caption{\raggedright 
   Physical properties of the critical TFIM from quantum simulations. (a) The energy is extrapolated to the zero-noise limit ($\lambda=0$) in close agreement with MPS benchmarks (green/orange lines at lower left signify MPS/Simulator result respectively).  
      (b)  The power law decay of longitudinal correlations $\langle Z_i Z_{i+d} \rangle$ with distance $d$ confirms that the hardware accurately captures the spatial structure of the state, despite cumulative gate noise. 
      (c) Pairwise entanglement of nearest neighbors is witnessed by the minimum eigenvalue $\lambda_\text{min}$ of the partially transposed density matrix; applying error mitigation successfully shifts hardware results below the zero-threshold to match the ideal ground state. 
(d) Quantum simulations with varying parameter choices capture the QPT through the crossover of the nearest-neighbor correlation $\langle Z_i Z_{j} \rangle$ and the average magnetization $\sum_i|\langle X_i\rangle|/N$ 
near $h=1.0$.  (e) $\lambda_\text{min}$ 
reaches its most negative point near the critical field ($h \approx 1.0$), signaling the regime of peak pairwise entanglement within the spin chain. 
}  
\label{fig:tfim}
\end{figure*}

From the TFIM experimental data in Fig.~\ref{fig:tfim}(a), we fit the total energy of the optimal ansatz state and obtain the ZNE value -25.37, which has accuracy of 99.53$\%$ with the numerical MPS value -25.49, for the state near criticality, $h=1.0$. 

We then measure all the relevant correlations for the ground state mentioned in Sec.~\ref{3b3}. 
Upon applying ZNE, we occasionally observed unstable extrapolations in which the inferred uncertainty was large compared with the mean value, reflecting the sensitivity of extrapolation-based error mitigation to statistical fluctuations and fit conditioning~\cite{Temme2017,Endo2018}. 
To mitigate the influence of these high-uncertainty estimates, we exploit the periodic boundary conditions and average the correlations over all translation-equivalent spin pairs at fixed separation $d$, as shown in Fig.~\ref{fig:tfim}(b) (also refer to Fig.~\ref{fig:tfim_zz},\ref{fig:tfim_20_heatmap_error} for details). 
We observe a characteristic algebraic decay of the longitudinal correlations
near the quantum critical point.  Fitting the averaged correlations to
$C(d)\sim d^{-\eta}$ gives
$\eta_{\rm fit}=0.212\pm0.002$, compared with the exact TFIM critical
exponent $\eta_{\rm exact}=1/4$ in the thermodynamic limit~\cite{Alicea2023,PFEUTY197079,Sachdev_2011}.
The measured exponent captures the expected power-law behavior, although it is
lower than the exact value, which may reflect finite-size effects.
Hardware results closely track the ideal MPS benchmarks, though the influence of gate noise becomes more pronounced at some separations, resulting in a slight divergence from the exact values. 
All other relevant local correlation functions are treated and mitigated in an analogous manner, and show similar agreement with the exact results.

Fig.~\ref{fig:tfim}(c) shows the $\lambda_\text{min}$ as a function of distance between the spins. We observe a negative eigenvalue for nearest-neighbor spins ($d=1$), indicating the presence of nearest-neighbor entanglement at the quantum phase transition. This behavior is consistently observed in both the hardware measurements and the exact results. The black dashed line demarcates the quantum correlations from the classical correlations.  Fig.~\ref{fig:tfim}(c) also highlights the impact of noise mitigation: the unmitigated hardware results (light blue), obtained averaging over the results in Fig.~\ref{fig:ppt_raw}(a), show a significant upward shift due to decoherence, which obscures the entanglement signature. 
Applying error mitigation techniques and averaging over spins at varying distances in Fig.~\ref{fig:ppt_raw}(c) achieves high accuracy with the ideal MPS ground state, effectively suppressing hardware-induced bias.

Despite the presence of long-range correlations at criticality as shown in Fig.~\ref{fig:tfim}(b), the PPT criterion yields negative eigenvalues only for nearest-neighbor sites in the hardware results and for nearest- and next-nearest neighbors in the numerical ground states. This indicates that two-spin entanglement does not persist across the chain, and that the long-range two-spin correlations of the state are classical in nature [as described by (\ref{separable rho})].  The quantum critical states are known to violate the area law of entanglement with logarithmic corrections in $N$ \cite{zaletel2011logarithmic}, which implies the presence of long-range entanglement that is not seen in our analysis. It appears to us that, for the Ising model, this long-range entanglement must appear only in irreducible multipartite correlations of three and more spins \cite{dur2000three}, which are not captured by the current two-spin analysis.   Analyzing the hierarchy of multipartite entanglement is an interesting future direction, but beyond our scope here.

\begin{figure*}[!htbp]  
    \centering  
    \includegraphics[width=1\linewidth]{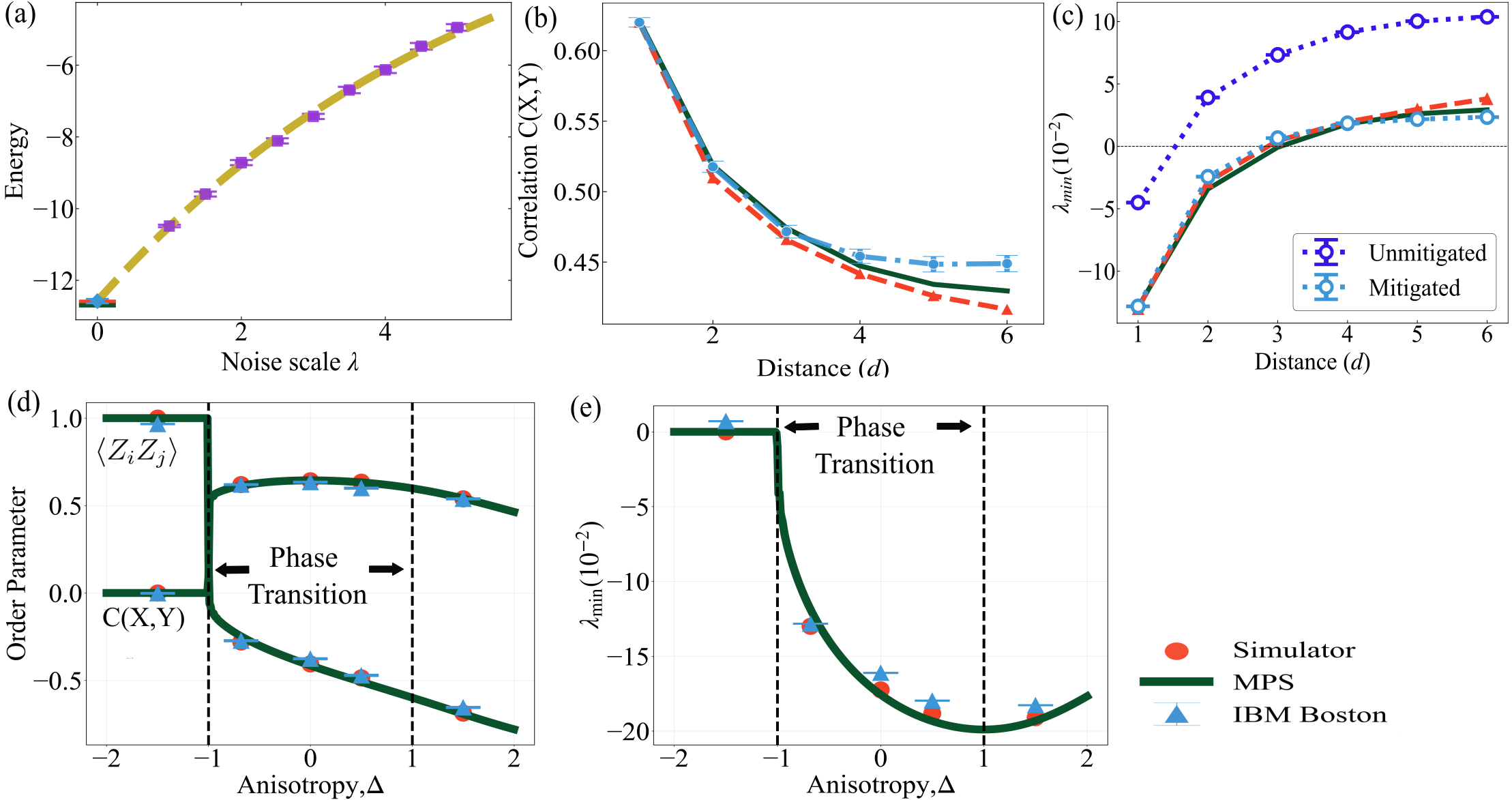}  
    
      \caption{\raggedright Quantum simulation results for the XXZ model analogous to Fig.~\ref{fig:tfim} (green/orange lines at lower left signify MPS/Simulator result respectively). (a) Energy estimation of the XXZ at $\Delta = -0.68$.
      (b) Long range transverse correlation  $C(X,Y)=(\langle X_i X_{i+d} \rangle + \langle Y_i Y_{i+d} \rangle)/2$ with distance $d$. 
(c) $\lambda_\text{min}$ witnesses entanglement   
 for nearest- and next-nearest neighbor pairs ($d=1,2$). 
(d) The change in $\langle Z_i Z_j\rangle$ and $C(X,Y)=\tfrac{1}{2}(\langle X_i X_j\rangle+\langle Y_i Y_j\rangle)$ near $\Delta=-1$ (and similarly near $\Delta=1$) marks the ferromagnetic–gapless and gapless–antiferromagnetic phase boundaries, respectively. 
(e) $\lambda_\text{min}$ within the gapless regime ($-1 < \Delta \leq 1$) exhibits the strongest quantum correlations. 
}  
\label{fig:xxz}
\end{figure*}

Next we examine how quantum simulations capture the approach to criticality by considering ground states prepared around the critical point. The order parameters $\langle Z_i Z_{i+1} \rangle$ and  $\sum_i |\langle X_i \rangle|/N$ change for states prepared around the phase transitions in Fig.~\ref{fig:tfim}(d). The characteristic crossing of the longitudinal correlations and transverse magnetization at $h \approx 1.0$ serves as a robust signature of the QPT, which is captured accurately in the quantum computations.  We used raw results and did not apply ZNE for shallow-depth circuits, since the inherent gate noise is sufficiently low that ZNE adds more uncertainty from fitting than any marginal accuracy gain. This applies in particular for transverse field values away from the critical point, at $h=0.01$ and $h=2.0$. 
\par 
The PPT criterion in Fig.~\ref{fig:tfim}(e) shows a clear dip near the  phase transition point. This dip marks the region where the states are most entangled, proving the presence of strong quantum correlations. Both the hardware (blue triangles) and the noiseless simulator results (orange circles) agree well with MPS simulations (dark green lines), validating the simulation framework and its ability to capture enhanced quantum correlations near quantum criticality. Additional results for $N=12$ are presented in Fig.~\ref{fig:tfim_12}.

\subsubsection{XXZ Model}\label{sec3c2}
We repeated all of the above analyses for the XXZ model. From the XXZ experimental data in Fig.~\ref{fig:xxz}(a), we obtain the ZNE value -12.58, which has accuracy of 99.05$\%$ with the numerical MPS value -12.70, for the state near QPT at $\Delta=-0.68$, again improving significantly on the unmitigated value.

We average the correlations over all equivalent spin separations $d$, as shown in Fig.~\ref{fig:xxz}(b) (refer to Fig.~\ref{fig:xxz_zz},~\ref{fig:xxz_12_heatmap_error} for more details), and again all the other local correlation functions are mitigated in a similar manner. The hardware results overall follows the general trend of the MPS results, capturing the long-range correlations of the gapless phase, but with a slight deviation at the largest separations $d=5,6$ due to cumulative gate noise.

Fig.~\ref{fig:xxz}(c) shows the PPT criterion as a function of distance between the spins near the critical point. The results reveal entanglement not only between nearest neighbors but also between next-nearest neighbors.
The agreement between the mitigated hardware data (dark blue), obtained averaging over the results in Fig.~\ref{fig:ppt_raw}(d), and the ideal MPS benchmark (dark green) demonstrates the robustness of error-mitigation in resolving the spatial extent of quantum correlations in the XXZ chain. While we encounter long range correlations in the gapless phase in Fig.~\ref{fig:xxz}(b), the PPT criterion shows that two-spin quantum correlations are absent beyond the next nearest neighbor pairs.

We observe the change in the local correlation functions $\langle Z_i Z_{i+1} \rangle$ and $C(X,Y)=\langle X_i X_{i+1} \rangle + \langle Y_i Y_{i+1} \rangle/2$ for parameter choices near the phase transition in Fig.~\ref{fig:xxz}(d). The transition at $\Delta = -1.0$ identifies the boundary between the gapped ferromagnetic and gapless phases, providing a distinct signature of the first-order phase transition. The PPT criterion shows the states in the gapless phase, $-1 < \Delta \leq 1$, are most entangled in Fig.~\ref{fig:xxz}(e), signifying the strongest quantum correlations.  Again, we used raw results and did not apply ZNE for shallow depth circuits, for $\Delta<-1$ and $\Delta>1$.  Additional results for $N=20$ are presented in Fig.~\ref{fig:xxz_20}.

\section{Conclusion}\label{sec7}

We systematically characterized two-qubit entanglement in quantum simualtions, using an efficient and model independent approach based on the PPT criterion.  We tested the approach by
preparing ground states of the TFIM and XXZ chains with MPS‑derived variational circuits and reconstructing all two‑qubit reduced density matrices via quantum overlapping tomography. Our results demonstrate that the PPT witness effectively identifies quantum phase transitions through clear peaks, indicating strong quantum correlations, and also provides a universal metric to benchmark quantum hardware. Furthermore, we have applied robust error-mitigation techniques---such as matrix-free readout correction and zero-noise extrapolation---obtaining quantitative agreement with MPS results on the IBMQ‑Boston device. The approach thus offers a model‑independent, near‑term‑compatible tool for probing many‑body entanglement and validating quantum computers, paving the way for systematic benchmarking of increasingly complex quantum simulations.

 To move toward a comprehensive understanding of many-body systems, an interesting future direction is to extend these benchmarks beyond two-spin entanglement to multipartite entanglement, which is essential for verifying the generation of genuinely large-scale quantum states~\cite{horodecki2009}. In contrast to standard experimental observables, such as neutron scattering or magnetization, which primarily probe local order or two-point correlations, multipartite extensions of PPT-based witnesses and overlapping tomography could reveal higher-order entanglement structures in quantum materials~\cite{liu2025,marsh2024}. Such techniques would provide a direct window into the fundamental entanglement structure of many-body systems, including regimes that are difficult to access with traditional condensed matter probes~\cite{Chen2022}. This capability could be particularly valuable in higher-dimensional systems, which remain challenging for classical high-performance computing and may offer a promising route toward demonstrating practical quantum utility~\cite{orus2013}. These entanglement witnesses and extensions of overlapping tomography could identify exotic phases such as quantum spin liquids, topological insulators, and superconductors, by detecting topological or non-local entanglement signatures rather than relying on local order parameters or broken symmetries~\cite{liu2025,lyu2025,bastian2011,alba2017}.

Another critical frontier involves studying non-equilibrium phenomena, for example in quantum quenches, where the behavior of time-evolved states can provide a rigorous stress test for gate fidelity and error-mitigation strategies~\cite{polkovnikov2011}. Characterizing the temporal spread of entanglement can serve as a dynamic benchmark for a processor’s capacity to sustain quantum advantage during execution~\cite{arute2019}. Finally, testing these entanglement witnesses across diverse architectures—such as superconducting, trapped-ion, and photonic systems—will facilitate a direct comparison of how different platforms handle the complexity of many-body entanglement.

\section*{Acknowledgments}\label{sec8}
A.B. and P.C.L. acknowledge funding from the US Department of Energy, the Office of Science through the Quantum Science Center (QSC), a National Quantum Information Science Research Center for project conception, design, execution, analysis, and writing. This research used resources of the Oak Ridge Leadership Computing Facility, which is a DOE Office of Science User Facility supported under Contract DE-AC05-00OR22725. This research used resources of the Compute and Data Environment for Science (CADES) at the Oak Ridge National Laboratory, which is supported by the Office of Science of the U.S. Department of Energy under Contract No. DE-AC05-00OR22725.

\bibliographystyle{unsrt}
\bibliography{ref}

\appendix
\section{State Preparation of the Model Hamiltonians}\label{appendix1}

To prepare the ground state of a Hamiltonian, $\hat{\mathrm{H}}$, we start with the ground state, $|\psi_\text{small} \rangle$ for the two-site Hamiltonian, $\hat{\mathrm{H}}_\text{small}$, 
using Density Matrix Renormalization Group (DMRG). 

DMRG is a powerful numerical method for computing low-lying states and observables of one-dimensional lattice spin models ~\cite{whitedmrg_92,whitedmrg_2023}. It is closely connected to matrix product states (MPS) ~\cite{ostulndmrg_1995,deukelskydmrg_1998,schollwockdmrg_2011}, also known as tensor trains ~\cite{dmrgbook_hackbush_2012}, which efficiently represent gapped 1D quantum systems with local interactions ~\cite{schollwockdmrg_2011,schollwockdmrg_2005,orusdmrg_2019,verstraetedmrg_2023}. 

We initially attempted to construct circuits directly from the MPS representation obtained from the DMRG results. However, this approach resulted in multi-qubit gates acting on many qubits at once, rather than the two-qubit gates that are typically supported by quantum hardware. To address this limitation, we instead adopted a brick-wall ansatz, which better matches the native two-qubit gate structure available on current quantum devices~\cite{yu2023}.

We decompose the nearest-neighbor Hamiltonian into two commuting parts,
$$\hat H = \hat H_{\mathrm{odd}} + \hat H_{\mathrm{even}},$$where $\hat H_{\mathrm{odd}}$ contains only interactions on odd bonds $(1,2),(3,4),\ldots,(N-1,N)$, and $\hat H_{\mathrm{even}}$ contains only interactions on even bonds $(2,3),(4,5),\ldots$. For even $N$, all terms within each layer act on disjoint pairs of sites. Starting with the ground state of the Hamiltonian, $\hat{\mathrm{H}}_\text{odd}$ 
having interaction only on odd numbers of bonds using $|\psi_\text{small} \rangle$. $\hat{\mathrm{H}}_\text{odd}$ has its unique ground state being the product of $|\psi_\text{small} \rangle$ over these odd bonds.  
\begin{equation}
    |\psi_\text{odd} \rangle = \Pi^{N/2}_{i=1}|\psi_\text{small} \rangle_{2i-1,2i}
\end{equation}
We expect that $|\psi_\text{odd} \rangle$ is adiabatically connected to the ground state of the total system, by connecting $\hat{\mathrm{H}}_\text{odd}$ to the full Hamiltonian $\hat{\mathrm{H}}$ via the linear interpolation,and $s$ is defined as $\frac{t}{T}$. 
\begin{equation}
        \hat{\mathrm{H}}(s)=(1-s)\hat{\mathrm{H}}_\text{odd}+s\hat{\mathrm{H}}=\hat{\mathrm{H}}_{o}(s)+\hat{\mathrm{H}}_{e}(s)
\end{equation}
We regroup it into interaction terms on even and odd bonds, denoted collectively by $\hat{\mathrm{H}}_{o}(s)$ and $\hat{\mathrm{H}}_{e}(s)$, respectively, and it is straightforward to see that $\hat{\mathrm{H}}_{o}(s)=\hat{\mathrm{H}}_\text{odd}$, but $\hat{\mathrm{H}}_{e}(s)$ is a rescaled version of the model on even bonds. $|\psi_\text{odd} \rangle$ is connected to the ground state of the $\hat{\mathrm{H}}$ via the evolution.
\begin{align}
        |\psi \rangle & =  U_{evo} |\psi_\text{odd} \rangle \nonumber \\
        & =  \exp({- \iota \mathcal{T}\int^{1}_{s=0}ds \hat{\mathrm{H}}(s)})|\psi_\text{odd} \rangle \nonumber \\ &= {\exp({- \iota \int^{T}_{t=0}\frac{dt}{ T}  \hat{\mathrm{H}}(\frac{ t}{ T})})}|\psi_\text{odd} \rangle
\end{align}
where $\iota$ is the imaginary number. Discretizing the evolution operator $\mathrm{U}_{evo}$, we have the following Trotterized approximation,

\begin{align}
    \mathrm{U}_{evo} & \simeq \Pi^{L}_{n=1}\exp(- \iota ds \hat{\mathrm{H}}(s)) \nonumber\\
    & \simeq \Pi^{L}_{n=1}\exp({- \iota ds \hat{\mathrm{H}}_{e}(s)})\exp({- \iota ds \hat{\mathrm{H}}_{o}(s)}) \nonumber\\
    & \simeq \Pi^{L}_{n=1}\exp\left({- \iota \frac{\Delta t}{ T} \hat{\mathrm{H}}_\text{o}}\right)       
        \exp\left({- \iota \frac{\Delta t}{ T}\hat{\mathrm{H}}_\text{e}}\right)
        \label{eq:U_evo_eqn}
\end{align}

where $L$ is the number of discretized time steps or layers and $\Delta s = 1/L$ is the dimensionless step size.
To allow for flexibility, the discretized evolution is turned into a variational form and arrive at the structure of ansatz, as shown in Fig~\ref{fig:loop_corr_ppt} (d)with gates in the $n$-th layer being $\mathrm{U}^{(n)}_\text{even/odd}({\theta})$. $\bm{\theta}'s$ are a set of variational parameters, with even and odd bonds $\{e/o\}$, respectively.

The notation e.g. $\theta^{n}_{e/o,z}$ depends only on the layer and the even/oddness of the qubit pairs.  It implies all even pairs in a layer have the same angle as each other, and all odd qubit pairs in a layer have the same angle as each other. We optimize $\Delta t$ first and fix it during the  evolution, 
and arrive at the following $L$-layer variational ansatz state in Eq. A5. We initialize the circuit parameters using a simple linear schedule (specified below for each model), and then optimize the individual parameters $\theta^n_{e/o}$ layer by layer until the fidelity is maximized.

    \begin{equation}
        |\psi_{ansatz}({\theta}) \rangle = \Pi^{L}_{n=1}[\mathrm{U}^{(n)}_\text{even}({\theta_{e}})\mathrm{U}^{(n)}_\text{odd}({\theta_{o}})] |\psi_\text{odd} \rangle
    \end{equation}
  Our goal is to maximize the fidelity between the ground-state wavefunction prepared by the quantum circuit and the ground state obtained from DMRG of the full quantum  Hamiltonian. Unlike QAOA, which typically minimizes the Hamiltonian’s energy, we directly optimize this wavefunction overlap.

\subsection{Creation of the ground state for the Hamiltonian $\hat{H}_\text{odd}$}\label{appendix1a}
A general MPS consisting of two qubits (two sites) as 
\begin{equation}
    |\phi\rangle= \sum_{a_1,a_2}\sum_{s_1,s_2}A^{[1]}_{s_1,a_1}A^{[2]}_{s_2,a_1 a_2}\Pi^{N=2}_{n=1}|s_n \rangle,
\end{equation}
with physical indexes  $s_n=\{\uparrow,\downarrow\}$ and virtual indexes $a_n=\{a'_n,a''_n\}=\{0,1\}$.  We use the left orthogonal form for the MPS. The $A^{[1]}$ satisfies the normalization condition, and $A^{[2]}$ satisfies the left orthogonal conditions, i.e.,
\begin{equation}
    \text{Tr}(A^{[1]} A^{[1] \star})= \sum_{s_1,a_1}A^{[1]}_{s_1,a_1}A^{[1]\star}_{a_1,s_1}=1
\end{equation}

\begin{equation}
    \sum_{s_2} A^{[2]}_{s_2,a'_1}A^{[2]\star}_{a"_1,s_2}= \mathbf{I}_{a'_1,a"_1}
    \label{eq:mpd}
\end{equation}
the above equation makes $A^{[2]}$ unitary.

We cast the $A^{[1]}$ and $A^{[2]}$ tensors into $G^{[1]}$ and $G^{[2]}$ unitary gates respectively to represent it in a quantum circuit, as a unitary matrix $\hat{\mathrm{U}}$ operating on $|00 \rangle$,  the MPS $|\phi \rangle$ can be written as 
\begin{align}
    |\phi\rangle & = \sum_{a_1}\sum_{s_1,s_2}A^{[1]}_{s_1,a_1}A^{[2]}_{s_2,a_1 }\Pi^{N=2}_{n=1}|s_n \rangle \nonumber \\ &=G^{[2]}G^{[1]} | 00 \rangle = \hat{\mathrm{U}} |00 \rangle,
\end{align}

\begin{figure}[H]  
    \centering  
    \includegraphics[width=1.0\linewidth]{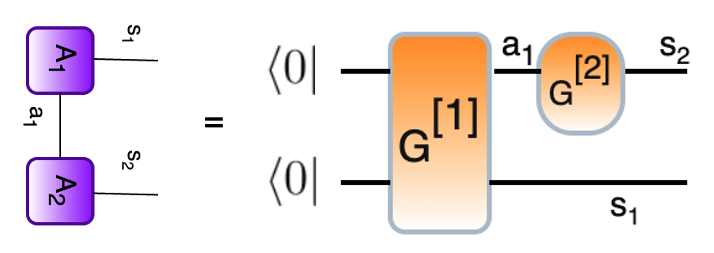}  
     
  \caption{ \raggedright Encoding a two qubit Matrix Product State exactly into a two qubit Quantum circuit.The MPS has two physical indices coming out $s_1$ and $s_2$, same as the quantum circuit.
}  
\label{fig:mps_qc}
\end{figure}

The bond index in the MPS $a_1$ comes in between the quantum circuit and then changes to the physical index $s_2$. The qubit change from a $|0 \rangle$ state to the bond virtual state $|a_1 \rangle$ and then to $|s_2 \rangle$ as shown in the Fig~\ref{fig:mps_qc}.
This procedure exactly maps the tensors in an MPS to unitary gates in a circuit. We get the two-qubit quantum state, $|\psi_\text{small} \rangle$, using this procedure.
\begin{equation}
    |\psi_\text{small} \rangle= \hat{\mathrm{U}}_\text{small} |00 \rangle
\end{equation}
We create the product of these two-qubit quantum state, $|\psi_\text{small} \rangle$ in parallel to represent the ground state of the hamiltonian, $\hat{\mathrm{H}}_\text{odd}$,
\begin{equation}
    |\psi_\text{odd} \rangle = \otimes^{N/2}_{i=1} |\psi_\text{small} \rangle =\otimes^{N/2}_{i=1} \hat{\mathrm{U}}_\text{small} |00 \rangle
\end{equation}

\subsection{State Preparation of the Transverse Field Ising Model}\label{appendix1b}

We implement the above procedure for the state preparation of $\hat{\mathrm{H}}_{\text{TFIM}}$, starting with the ground state of the Hamiltonian, $\hat{\mathrm{H}}_\text{odd}$ 

\begin{equation}
        \hat{\mathrm{H}}_\text{odd}=-J\sum^{N/2}_{i=1}\sigma^{z}_{2i-1}\sigma^{z}_{2i}+h/2\sum^{N/2}_{i=1}(\sigma^{x}_{2i-1}+\sigma^{x}_{2i}).
\end{equation}

 The essential two-qubit $R^{e/o}_{z,x}$ gate that we need for the TFIM, is related to the interaction terms of the Hamiltonian, $\hat{\mathrm{H}}_{\text{TFIM}}$,
\begin{equation}
    R^{e/o}_{z,x}(\theta_z,\theta_x)= e^{-\iota (\theta_z/2 )\sigma^z \otimes \sigma^z}e^{-\iota (\theta_x/2) \sigma^x}
\end{equation}
where a factor of 1/2 is inserted in the definition of the $R^{e/o}_{z,x}$ gate to match the convention of single-qubit rotation.

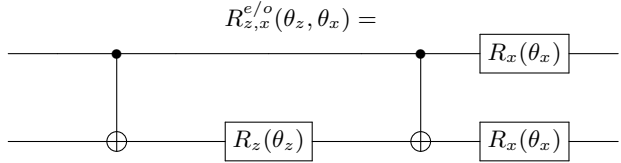
\begin{figure}[H]
\centering
$R^{e/o}_{z,x}(\theta_z,\theta_x) =
\begin{array}{c}
\Qcircuit @C=2em @R=2em {
 & \qw            & \ctrl{1}         & \qw                     & \qw                          & \qw           & \ctrl{1}        & \gate{R_{x}(\theta_x)} & \qw \\
 & \qw            & \targ        & \qw     & \gate{R_{z}(\theta_z)}   & \qw & \targ & \gate{R_{x}(\theta_x)} & \qw               
}
\end{array}
$
\caption{Ansatz circuit for TFIM.}
\label{fig:tfim_circ}
\end{figure}

We present a decomposition of the $R^{e/o}_{z,x}$ gate that has a minimum number of CNOTs( which is two), $R^{\alpha}(\theta)=e^{-\iota \theta \sigma^{\alpha}/2}$ is the single qubit rotation around $\alpha$-axis ($\alpha= x,y,z$) by an angle $\theta$. The circuit shown in Fig~\ref{fig:tfim_circ}
is symmetric with respect to swapping the two qubits. $\theta^{n}_{o,z}=(1- \Delta t/T)(-J)  \Delta t/T+( \Delta t/T)(-J) \Delta t/T, \theta^{n}_{e,z}=( \Delta t/T)(-J)  \Delta t/T, \theta^{n}_{o,x}=(1- \Delta t/T)(h)  \Delta t/T +(\Delta t/T)(h)  \Delta t/T,   \theta^{n}_{e,x}=( \Delta t/T)(h)  \Delta t/T$ are assigned as the initial parameters for the ansatz circuit, before being optimized layer-by-layer for the final circuits, and the angles are bounded between $\{ -\pi, \pi \}$ during optimization.

\subsection{State Preparation of the XXZ Model}\label{appendix1c}

The state preparation of $\hat{\mathrm{H}}_{\text{XXZ}}$ again follows the same procedure, starting with the ground state of the Hamiltonian, $\hat{\mathrm{H}}_\text{odd}$

\begin{equation}
        \hat{\mathrm{H}}_\text{odd}=-J\sum^{N/2-1}_{i=1}(\sigma^{x}_{2i-1}\sigma^{x}_{2i}+\sigma^{y}_{2i-1}\sigma^{y}_{2i})+ \Delta \sigma^{z}_{2i-1}\sigma^{z}_{2i}.
\end{equation}

The two-qubit $R^{e/o}_{x,y,z}$ gate for the XXZ model,
\begin{small}
\begin{equation}
    R^{e/o}_{x,y,z}(\theta_z,\theta_x)= e^{-\iota (\theta_x/2 )\sigma^x \otimes \sigma^x}e^{-\iota (\theta_y/2 )\sigma^y \otimes \sigma^y}e^{-\iota (\theta_z/2 )\sigma^z \otimes \sigma^z}
\end{equation}
\end{small}

\begin{figure*}[!htbp]

\centering
$R^{e/o}_{x,y,z}(\theta_x,\theta_y, \theta_z) =
\begin{array}{c}
\Qcircuit @C=0.5em @R=0.5em {
 & \qw & \targ      & \qw & \gate{R_{z}(\theta_z)} & \qw & \qw & \qw & \targ & \qw & \gate{R_{z}(-\theta_y)} & \qw & \targ & \qw & \gate{R_{x}(\pi/2)} & \qw \\
 & \qw & \ctrl{-1}  & \qw & \gate{H}               & \qw & \gate{R_{z}(\theta_x+\pi/2)} & \qw & \ctrl{-1} & \qw & \gate{H} & \qw & \ctrl{-1} & \qw & \gate{R_{x}(-\pi/2)} & \qw
}
\end{array}$
\caption{Ansatz circuit for XXZ.}
\label{fig:xxz_circ}
\end{figure*}
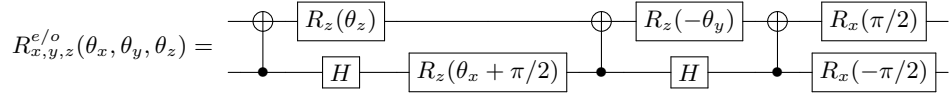

The circuit shown in Fig~\ref{fig:xxz_circ}
is symmetric. 
The initial parameters for the ansatz circuit are assigned analogously to the TFIM model, and the angles are bounded between $\{ -\pi, \pi \}$ during optimization. We use the BFGS method for the numerical optimization for both the Hamiltonians.

\begin{figure*}[!htbp]
\centering

\begin{subfigure}{0.45\textwidth}
\centering
\includegraphics[width=\linewidth]{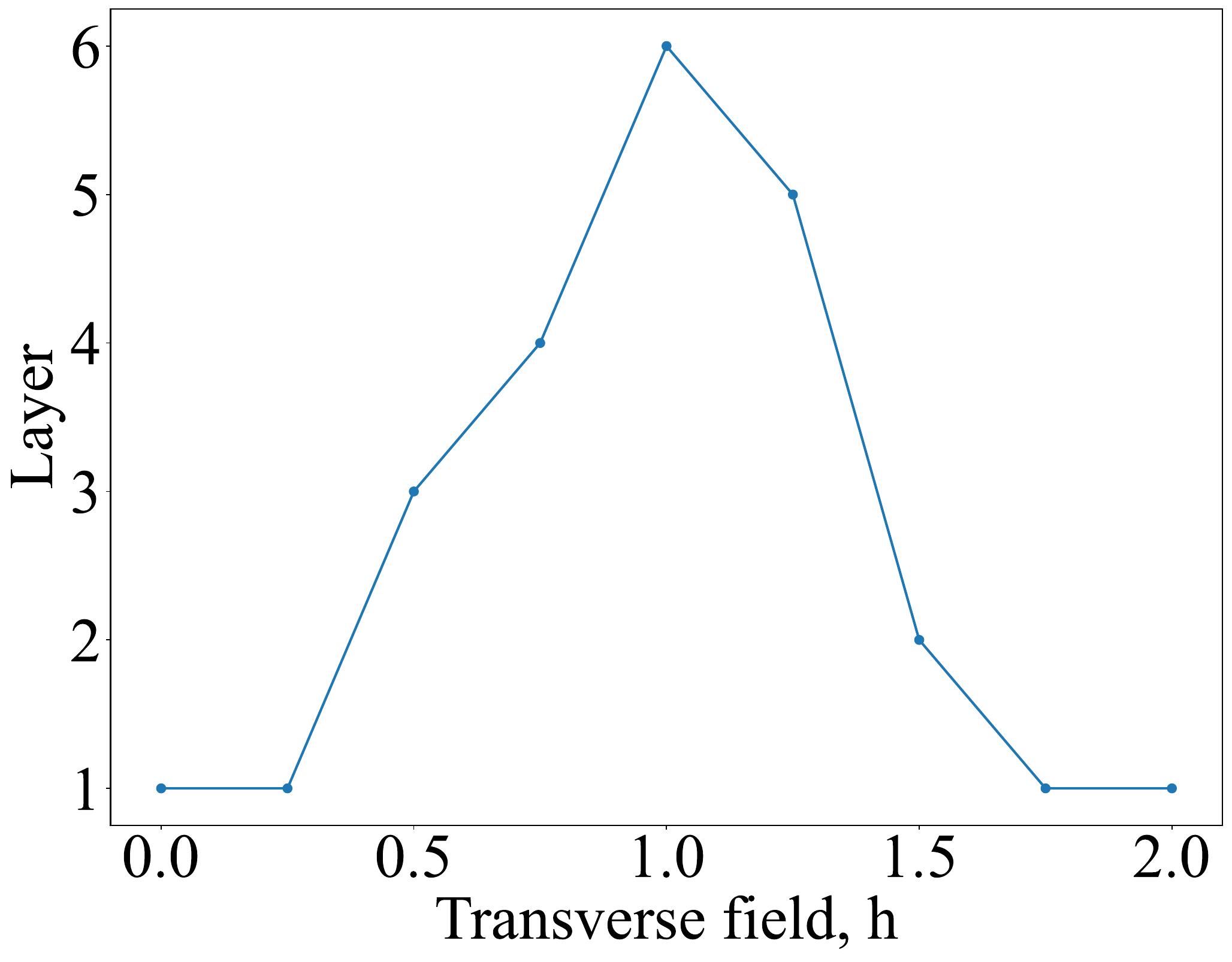}
\caption{}
\end{subfigure}
\hfill
\begin{subfigure}{0.48\textwidth}
\centering
\includegraphics[width=\linewidth]{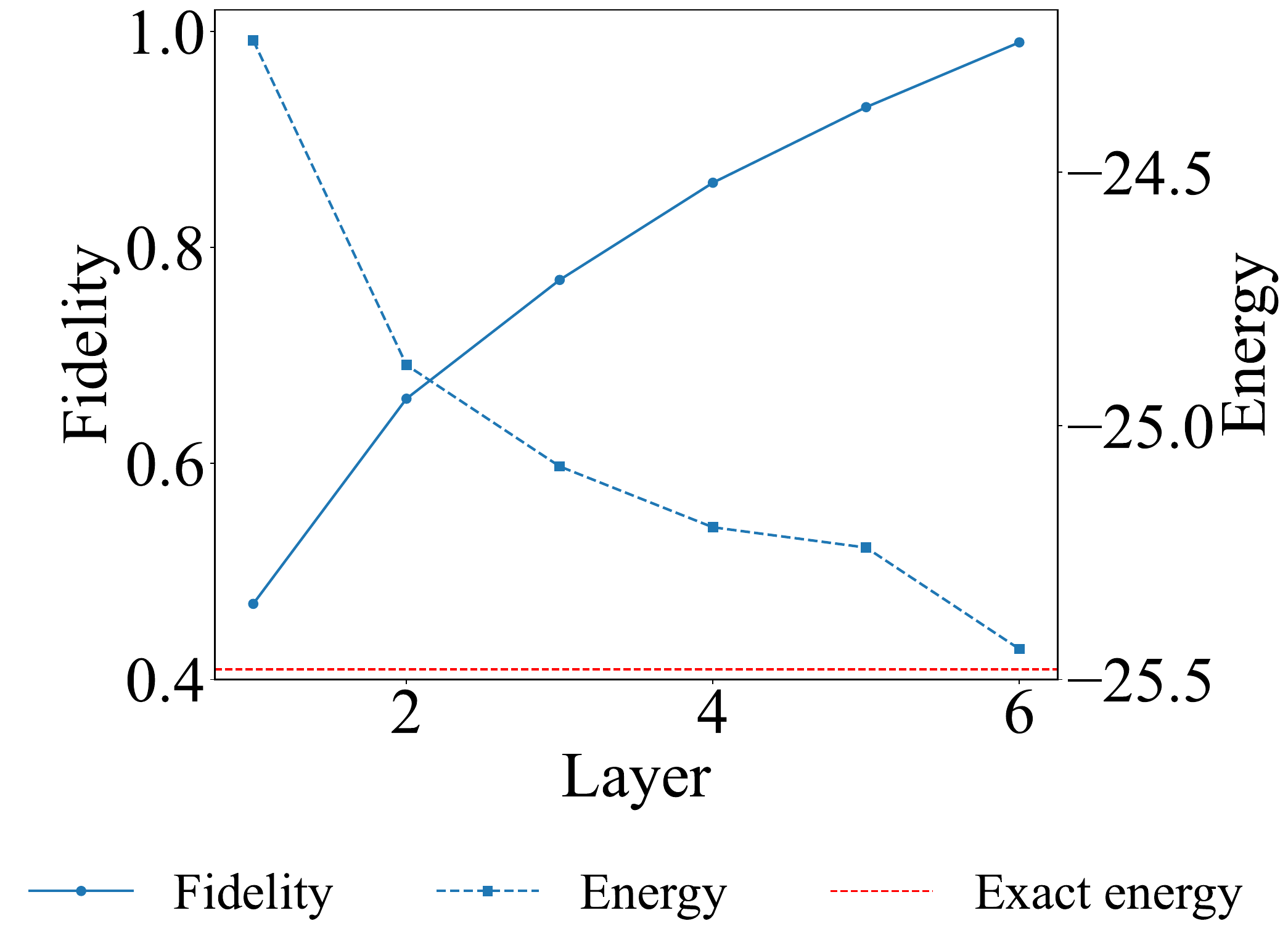}
\caption{}
\end{subfigure}

\vspace{0.3cm}

\begin{subfigure}{0.45\textwidth}
\centering
\includegraphics[width=\linewidth]{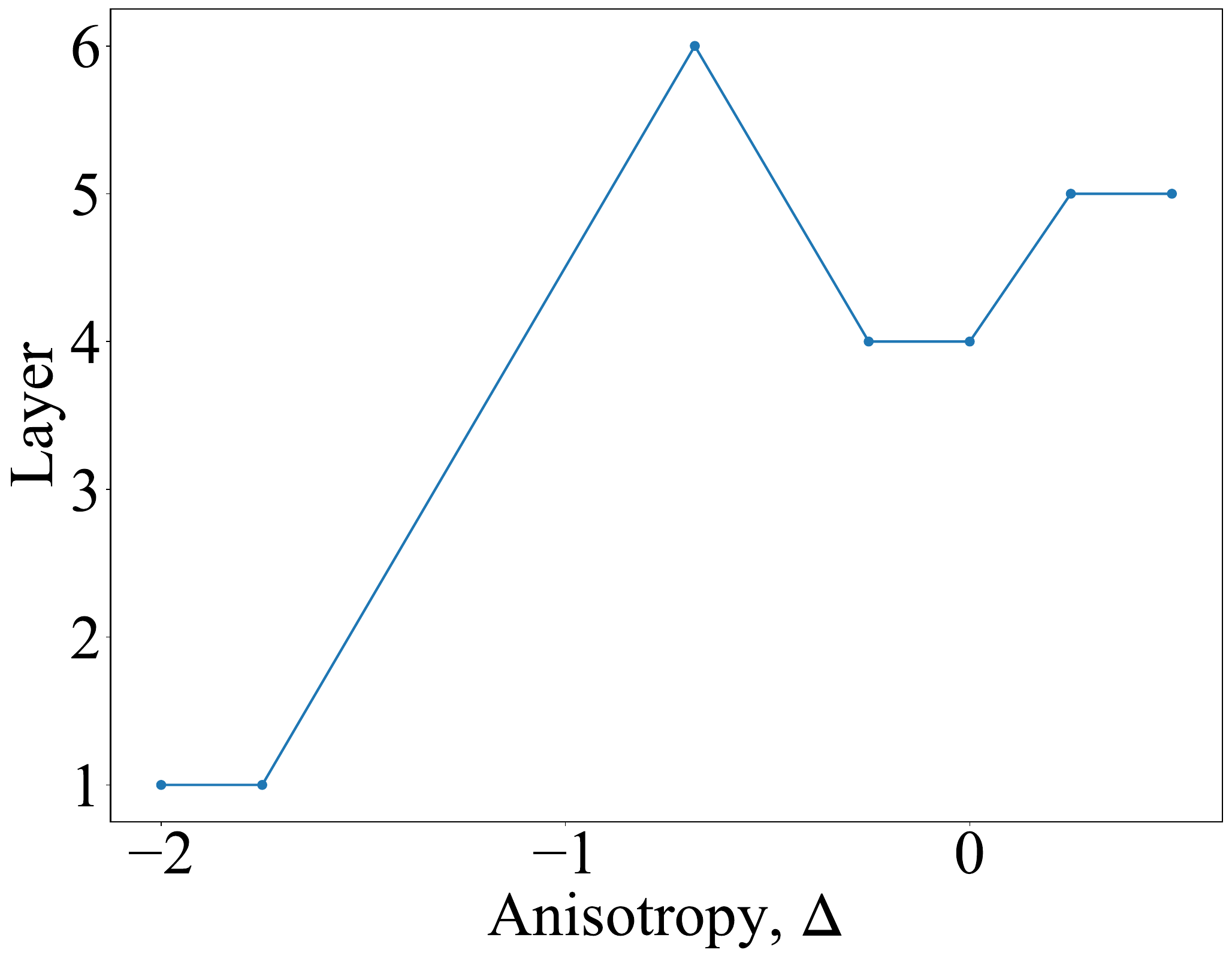}
\caption{}
\end{subfigure}
\hfill
\begin{subfigure}{0.48\textwidth}
\centering
\includegraphics[width=\linewidth]{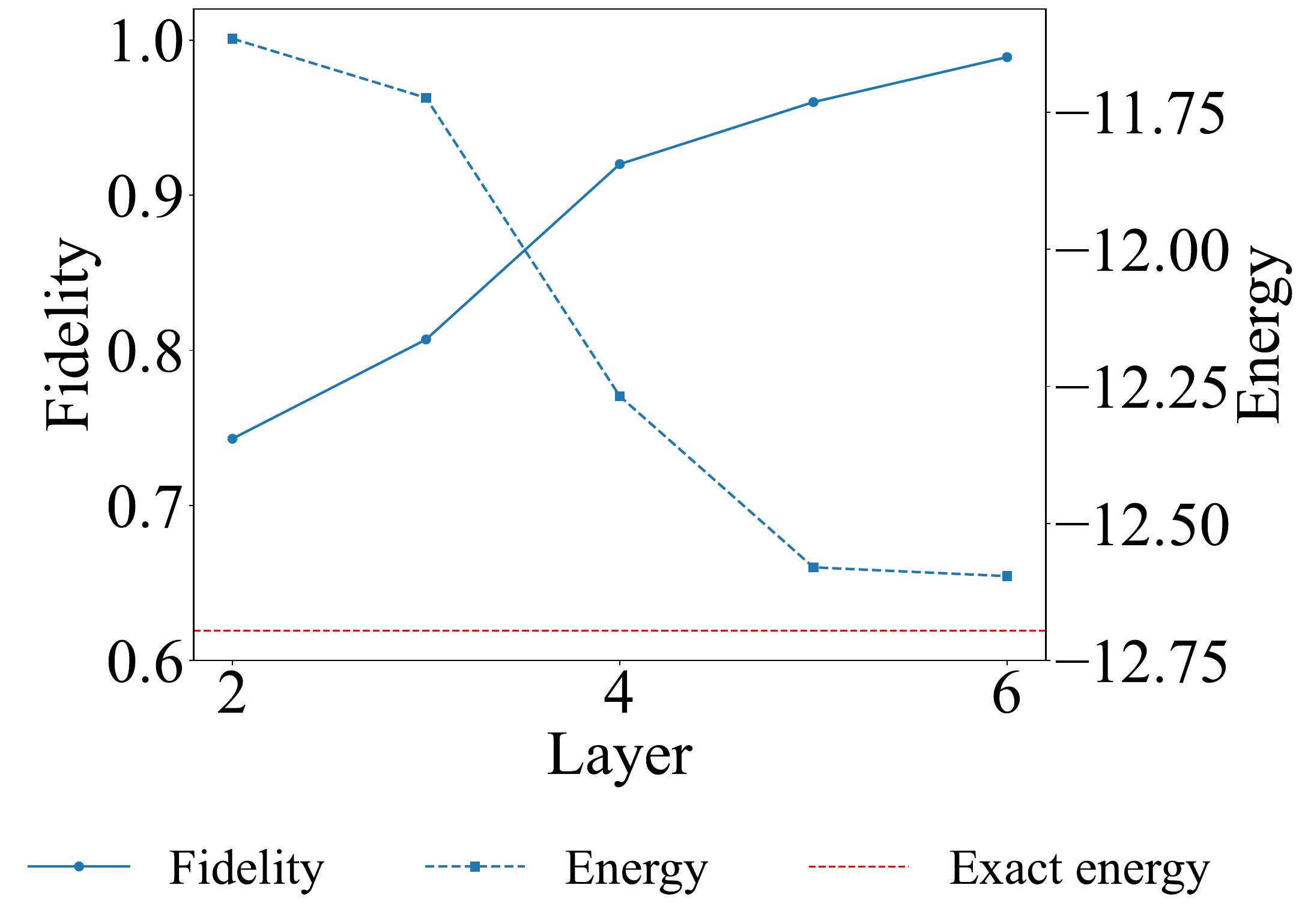}
\caption{}
\end{subfigure}

\caption{\raggedright State preparation of the spin models.
(a) Number of circuit layers required to optimize the fidelity between the ansatz state and the ground state of the TFIM, $N=20$, as a function of the transverse field strength $h$, and it increases near the quantum phase transition.(b) Fidelity and energy of the optimal ansatz state relative to the critical ground state as a function of the number of layers for the TFIM. The red line denotes the exact ground-state energy. (c,d) show similar results for the XXZ, $N=12$ model.}

\label{fig:layer_fid_energy}

\end{figure*}

\begin{figure*}[!htbp]
\centering

\begin{subfigure}{0.9\textwidth}
\centering
\includegraphics[width=\linewidth]{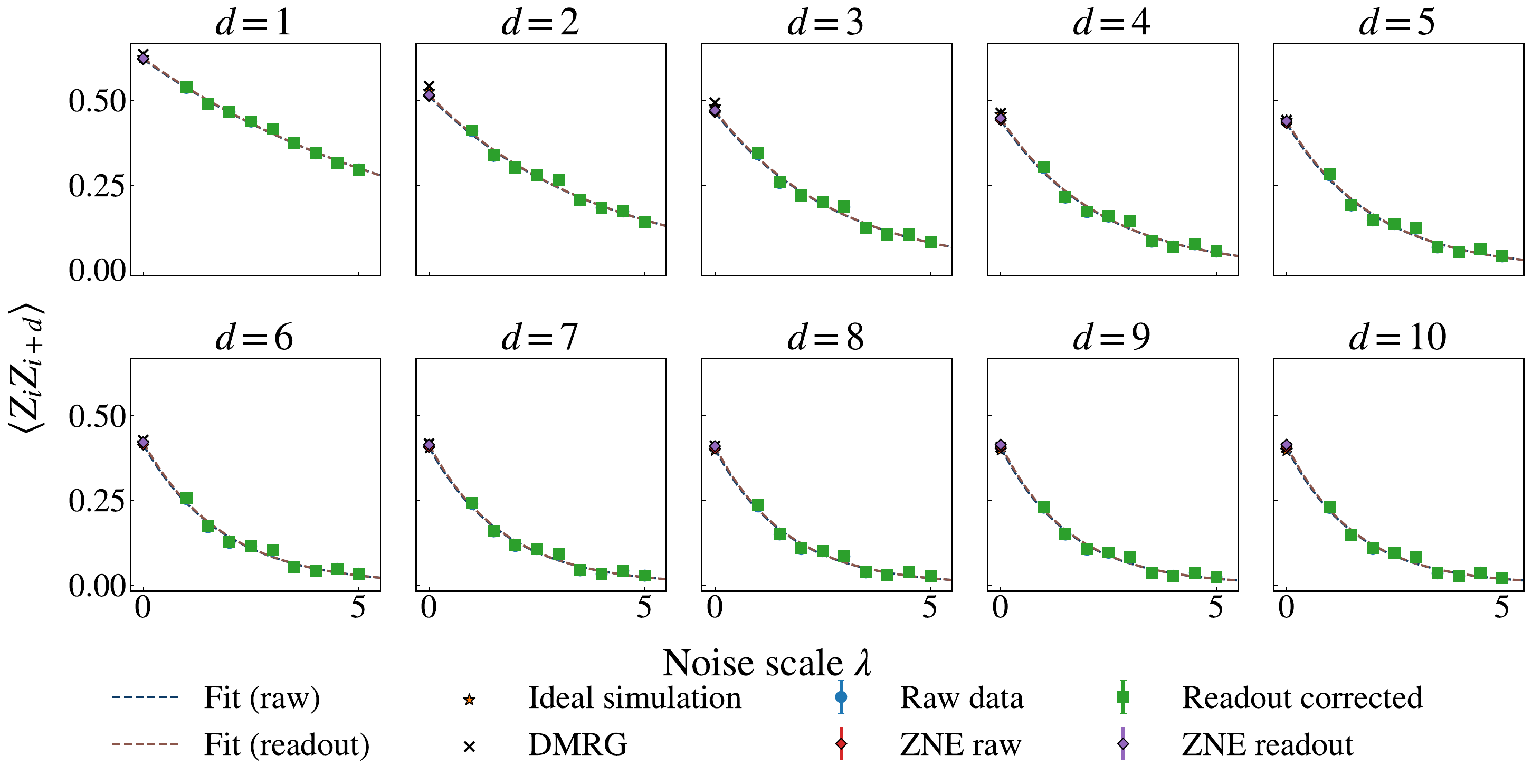}
\caption{}
\end{subfigure}

\vspace{0.6cm}

\begin{subfigure}{0.45\textwidth}
\centering
\includegraphics[width=\linewidth]{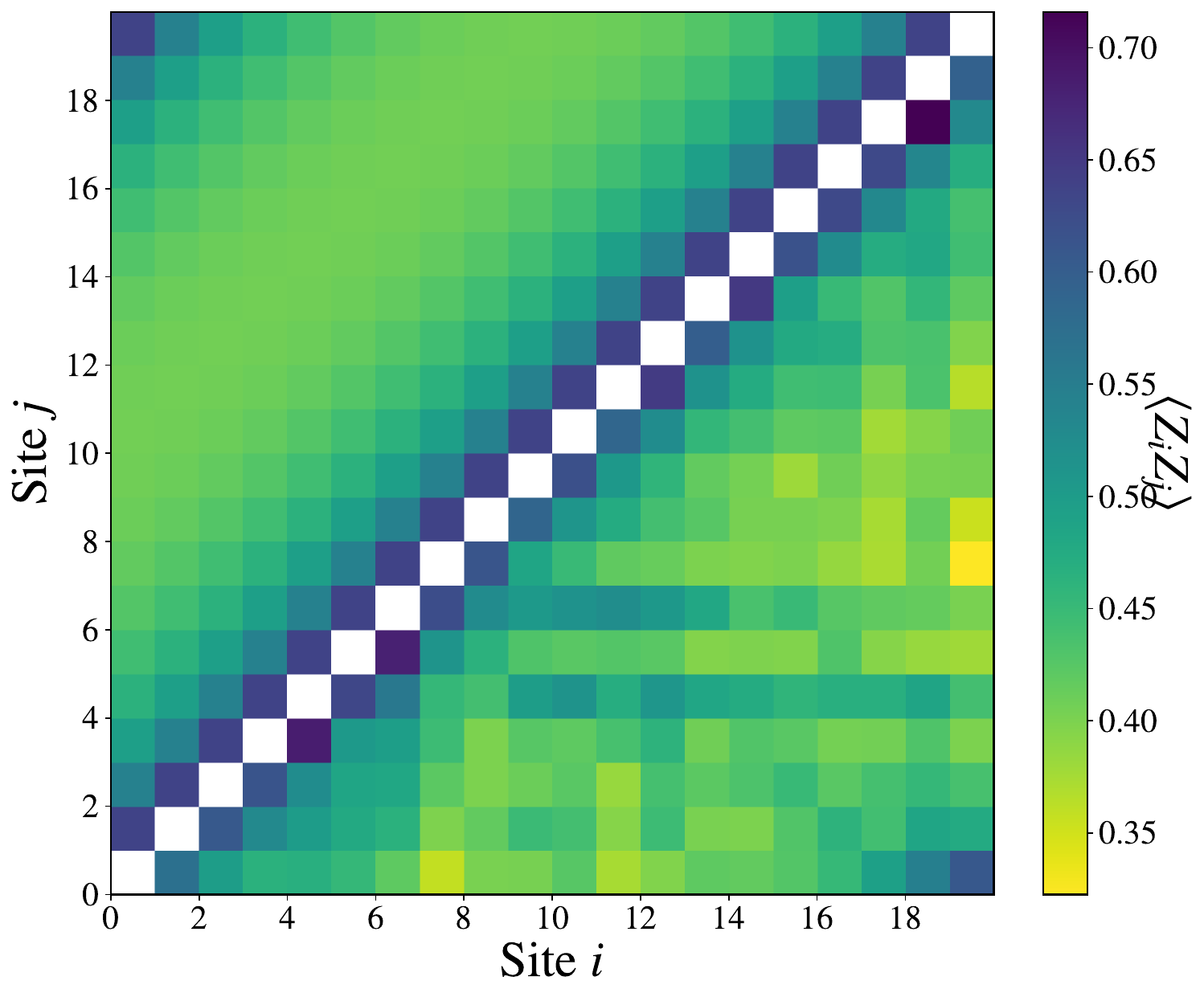}
\caption{}
\end{subfigure}
\hfill
\begin{subfigure}{0.45\textwidth}
\centering
\includegraphics[width=\linewidth]{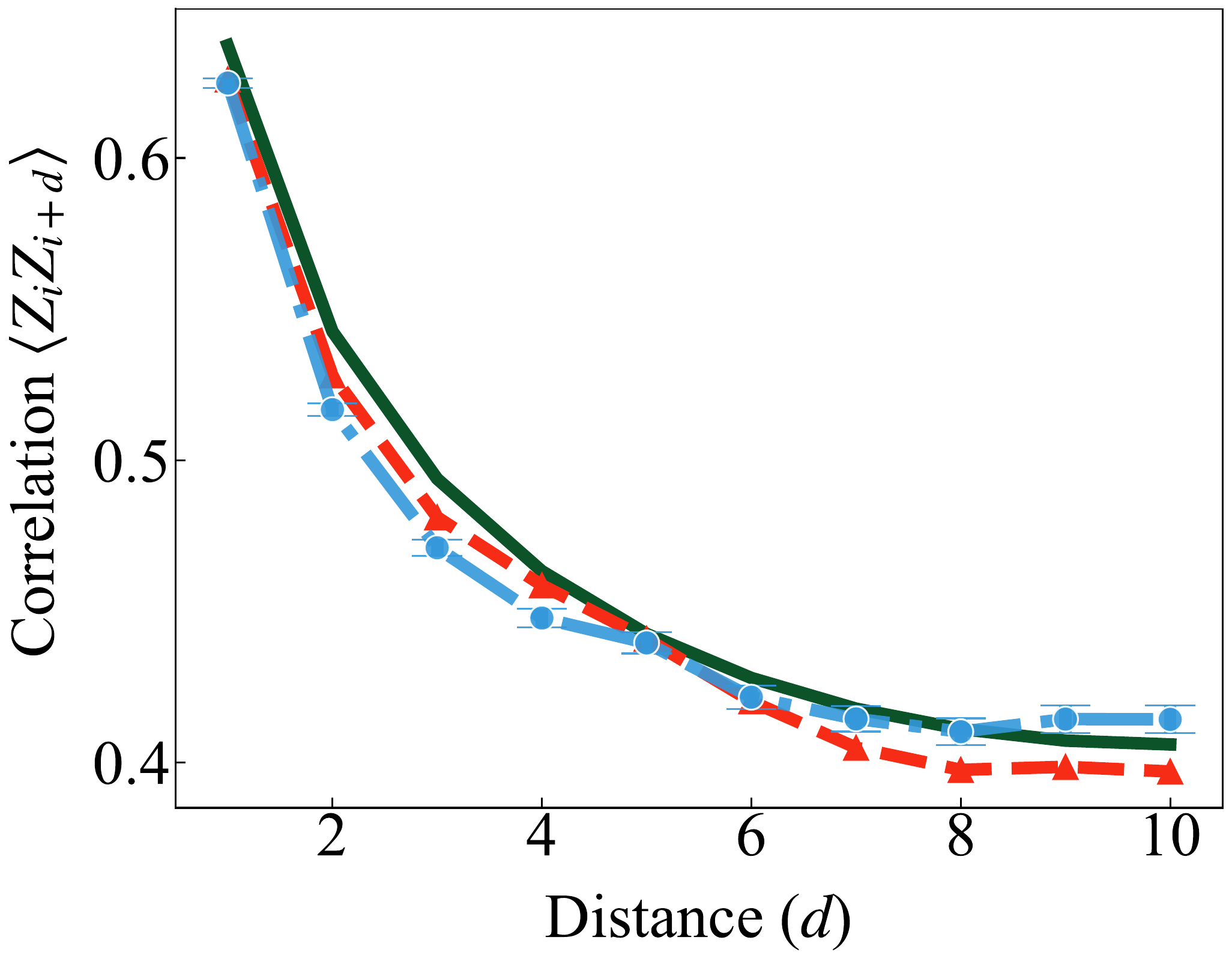}
\caption{}
\end{subfigure}

\caption{\raggedright
(a) Zero-noise extrapolation applied to cloud-based experimental measurements of the correlation function $\langle Z_i Z_{i+d} \rangle$ for different spin separations $d$ in the transverse-field Ising model (TFIM) at the critical field $h = 1.0$.(b) Correlation matrix $\langle Z_i Z_j \rangle$ between spins $i$ and $j$. The upper-left triangle ($i > j$) shows the exact results obtained from density matrix renormalization group (DMRG) calculations, while the lower-right triangle ($i < j$) displays the corresponding results measured on quantum hardware.(c) Zero-noise extrapolation for cloud-based experimental measurements of $\langle Z_i Z_{i+d} \rangle$, averaged over spin separation $d$, for the TFIM at $h = 1.0$. The reported data have already been corrected using measurement-error mitigation techniques.
}

\label{fig:tfim_zz}

\end{figure*}

\begin{figure*}[!htbp] 
\centering

\begin{subfigure}{0.9\textwidth}
\centering
\includegraphics[width=\linewidth]{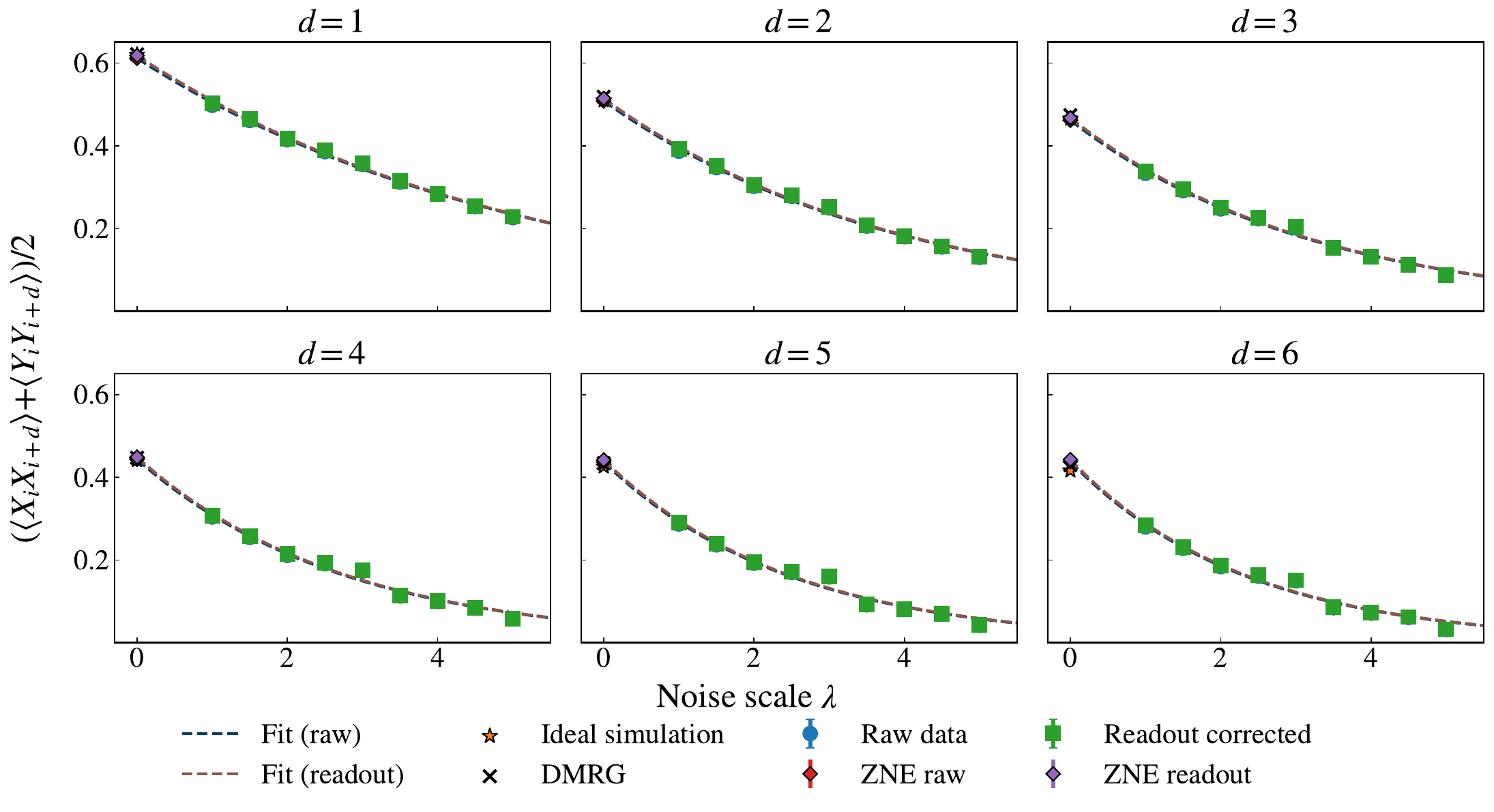}
\caption{}
\end{subfigure}

\vspace{0.6cm}

\begin{subfigure}{0.45\textwidth}
\centering
\includegraphics[width=\linewidth]{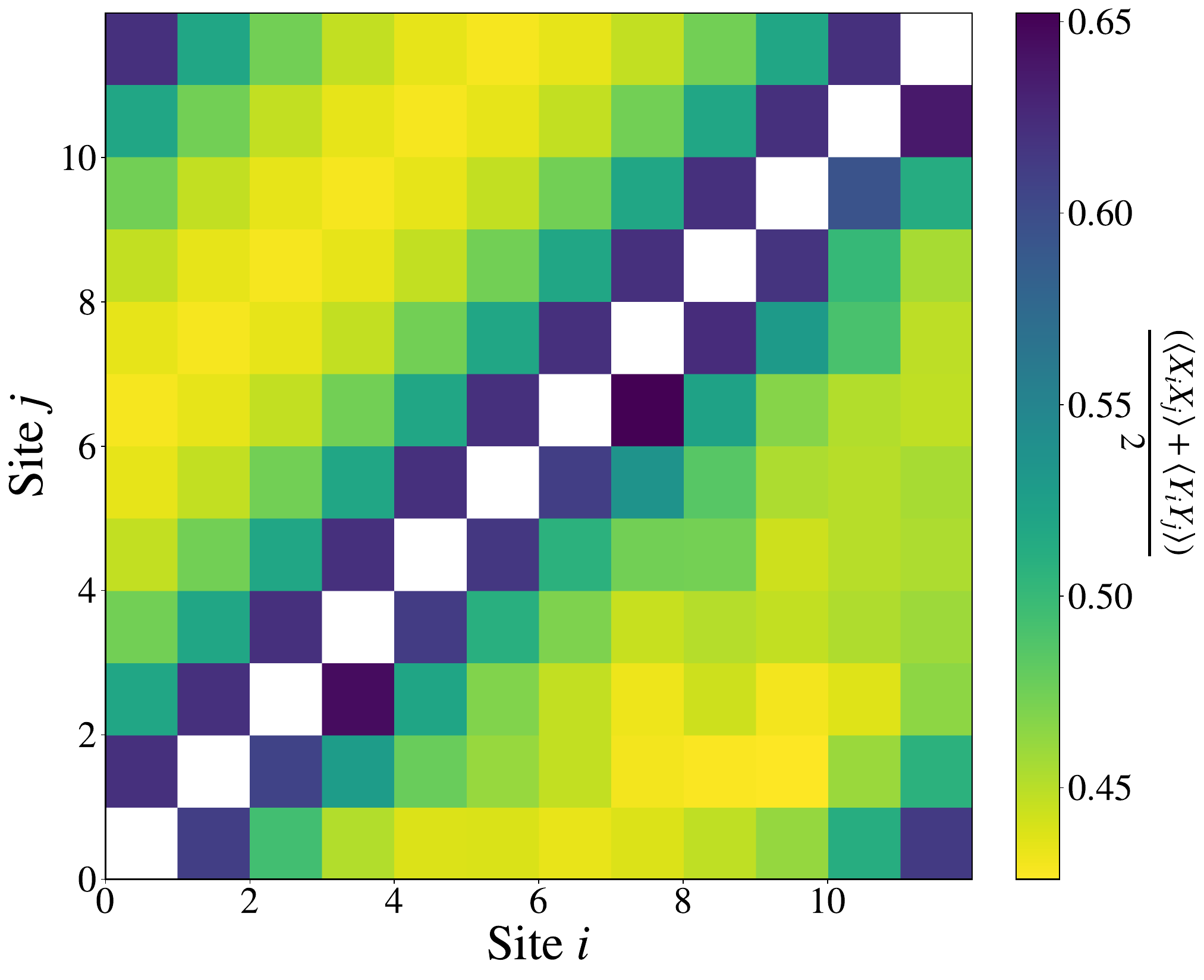}
\caption{}
\end{subfigure}
\hfill
\begin{subfigure}{0.45\textwidth}
\centering
\includegraphics[width=\linewidth]{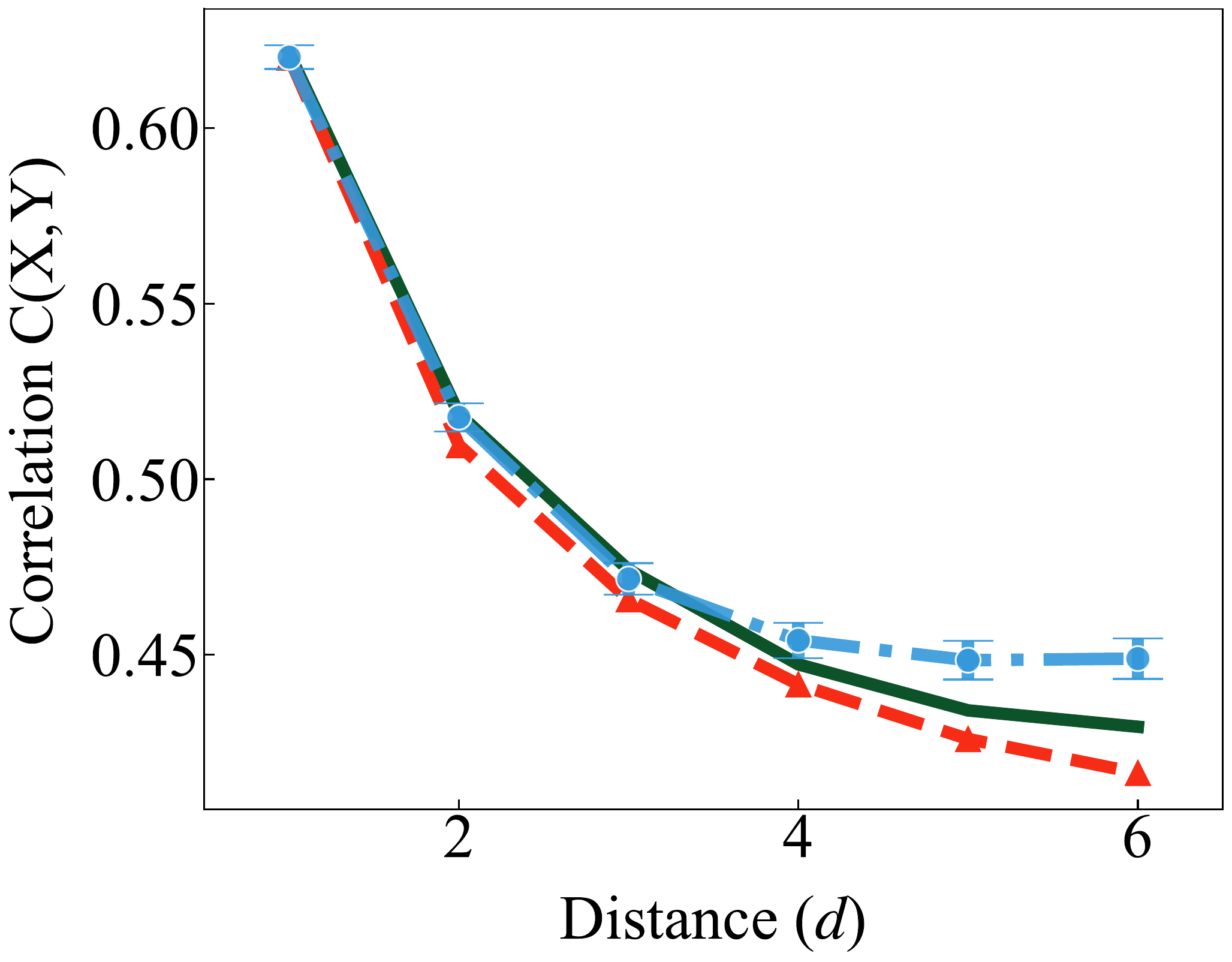}
\caption{}
\end{subfigure}

\caption{\raggedright (a) Zero-noise extrapolation applied to cloud-based experimental measurements of the correlation function $C(X,Y)=(\langle X_i X_{i+d} \rangle + \langle Y_i Y_{i+d} \rangle)/2.0$ for different spin separations $d$ in the XXZ model at the critical anisotropy $\Delta=-0.68$.(b)Correlation matrix $C(X,Y)=(\langle X_i X_{j} \rangle + \langle Y_i Y_{j} \rangle)/2.0$ between spins $i$ and $j$. The upper-left triangle ($i > j$) shows the exact results obtained from density matrix renormalization group (DMRG) calculations, while the lower-right triangle ($i < j$) shows the corresponding results measured on quantum hardware.(c)Zero-noise extrapolation for the cloud-based experimental measurements of $C(X,Y)=(\langle X_i X_{i+d} \rangle + \langle Y_i Y_{i+d} \rangle)/2.0$ averaged over spin separation $d$, for the XXZ model at $\Delta = -0.68$. The reported data have already been corrected using measurement-error mitigation techniques.}
\label{fig:xxz_zz}
\end{figure*}

\begin{figure*}[!htbp]
\centering
\begin{subfigure}{0.46\textwidth}
\centering
\includegraphics[width=\linewidth]{zz_heat_tfim_20.pdf}
\caption{}
\end{subfigure}
\hfill
\begin{subfigure}{0.47\textwidth}
\centering
\includegraphics[width=\linewidth]{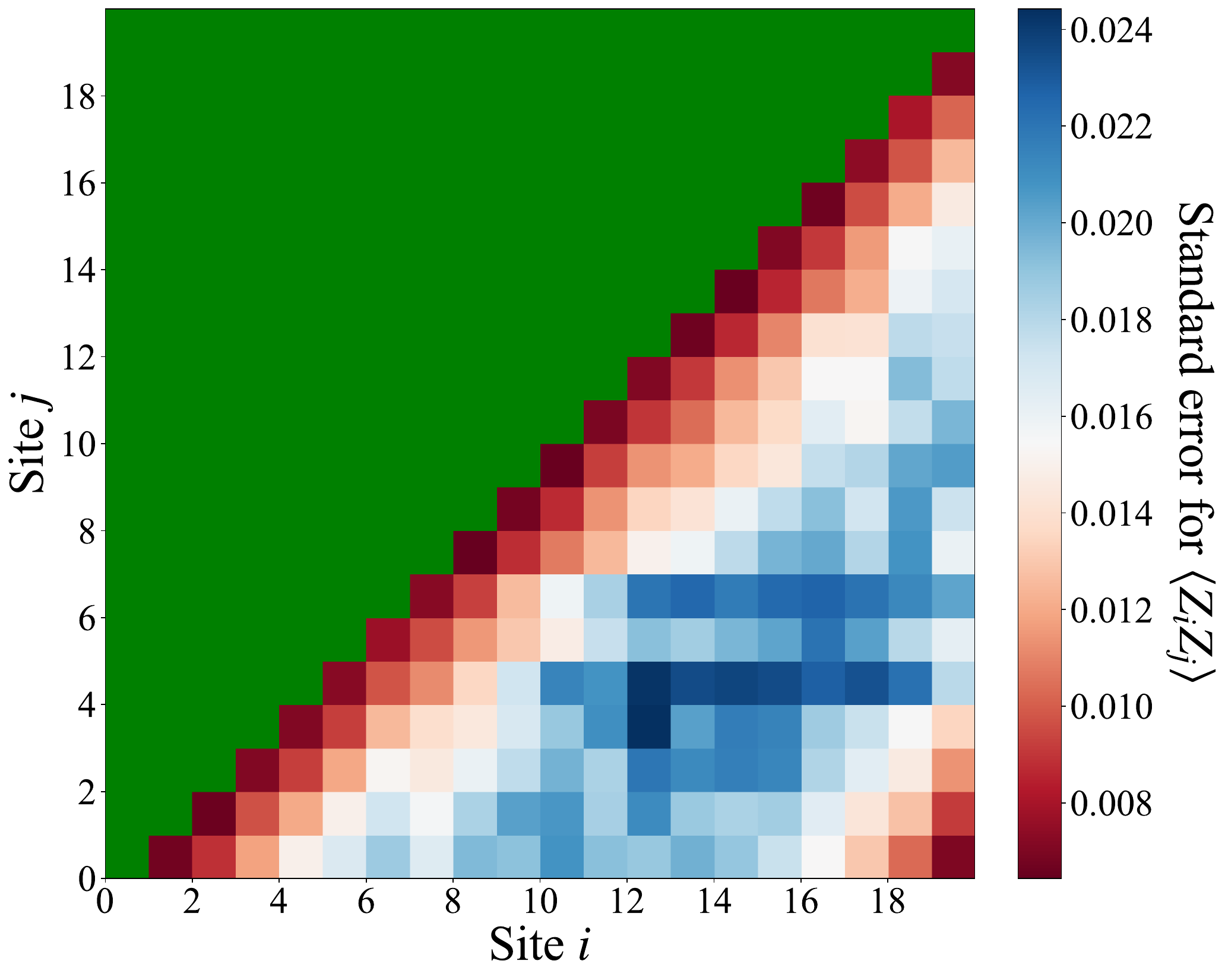}
\caption{}
\end{subfigure}
\vspace{0.3cm}
\begin{subfigure}{0.48\textwidth}
\centering
\includegraphics[width=\linewidth]{lmin_heat_tfim_raw_20.pdf}
\caption{}
\end{subfigure}
\hfill
\begin{subfigure}{0.47\textwidth}
\centering
\includegraphics[width=\linewidth]{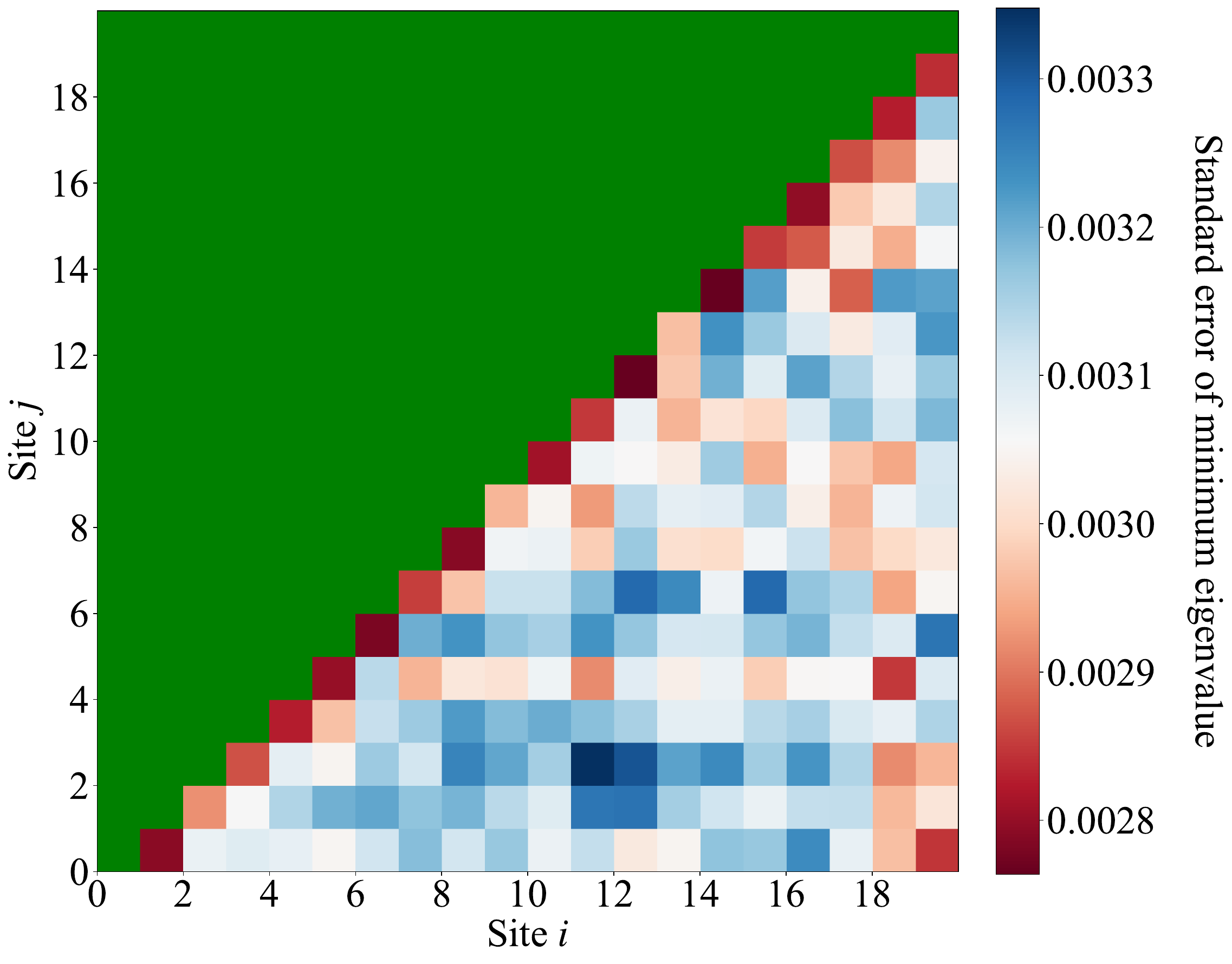}
\caption{}
\end{subfigure}
\vspace{0.3cm}
\begin{subfigure}{0.48\textwidth}
\centering
\includegraphics[width=\linewidth]{lmin_heat_tfim_miti_20.pdf}
\caption{}
\end{subfigure}
\hfill
\begin{subfigure}{0.47\textwidth}
\centering
\includegraphics[width=\linewidth]{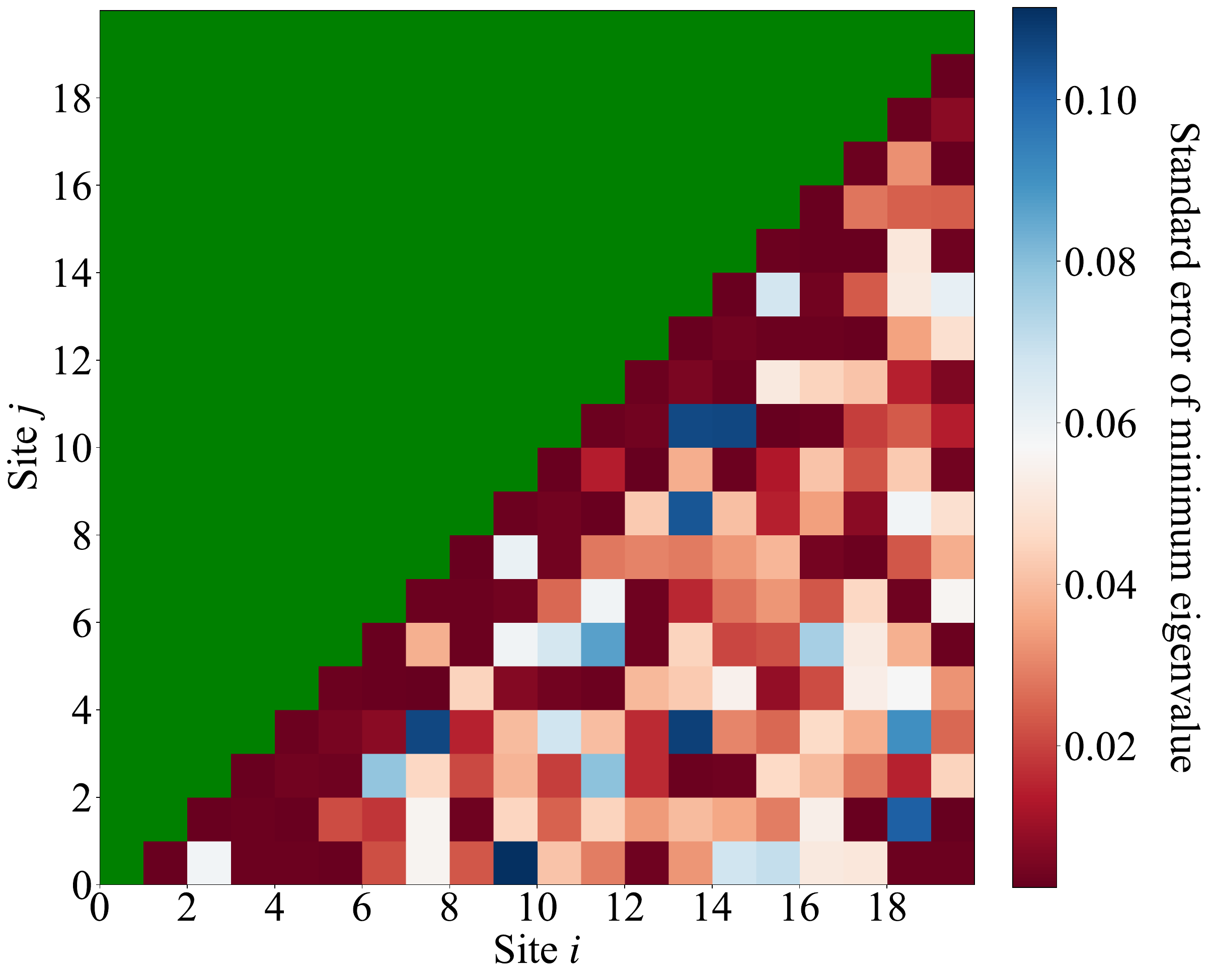}
\caption{}
\end{subfigure}
\caption{ Heatmaps (a,c,e) and corresponding error maps (b,d,f) for $N=20$ TFIM model. (a,b) Correlation function $\langle Z_i Z_{i+d} \rangle$. (c,d) Unmitigated $\lambda_\text{min}$. (e,f) $\lambda_\text{min}$ after error mitigation. }
\label{fig:tfim_20_heatmap_error}
\end{figure*}

\begin{figure*}[!htbp]
\centering
\begin{subfigure}{0.46\textwidth}
\centering
\includegraphics[width=\linewidth]{xx_yy_heat_xxz_12.pdf}
\caption{}
\end{subfigure}
\hfill
\begin{subfigure}{0.47\textwidth}
\centering
\includegraphics[width=\linewidth]{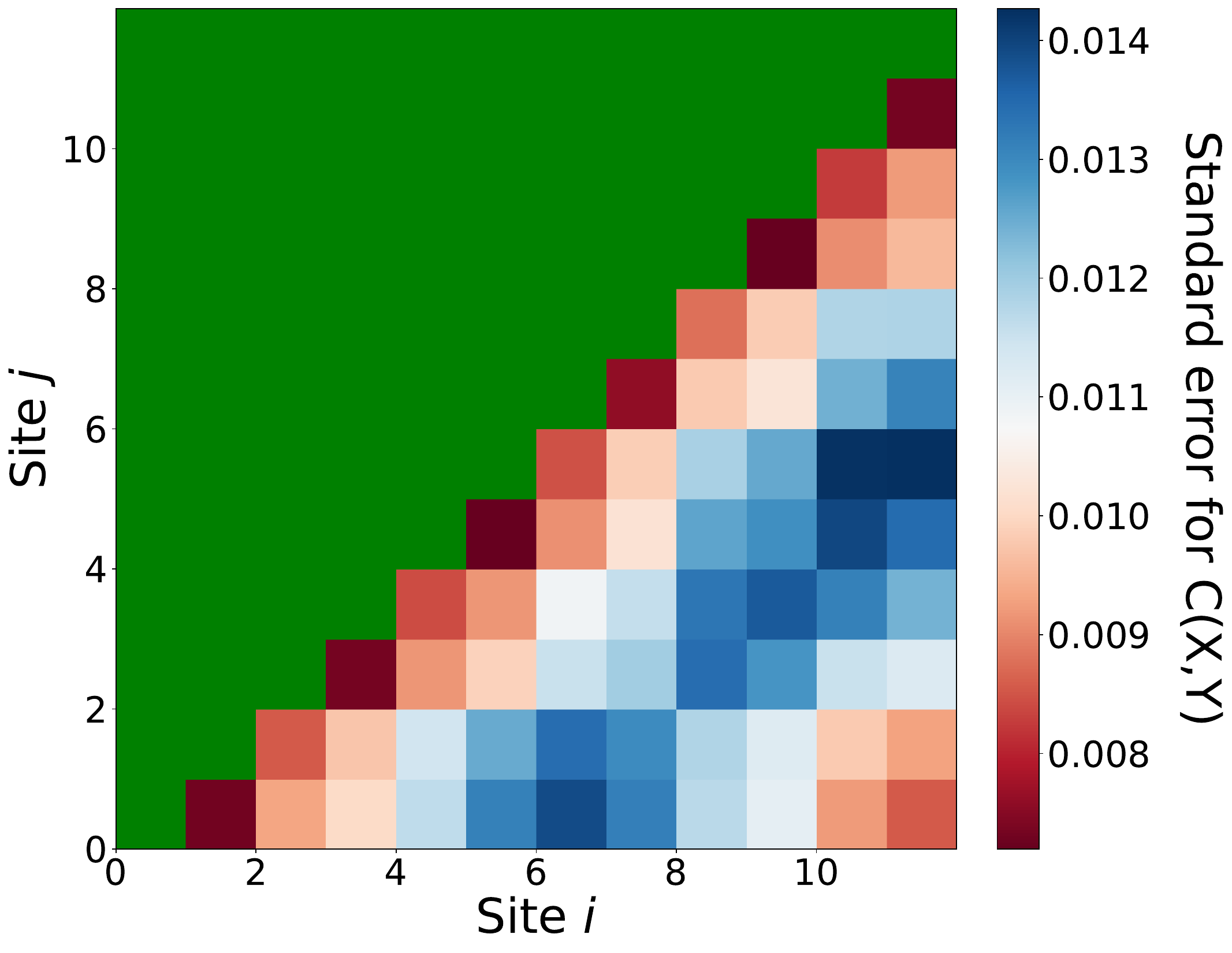}
\caption{}
\end{subfigure}
\vspace{0.3cm}
\begin{subfigure}{0.48\textwidth}
\centering
\includegraphics[width=\linewidth]{lmin_heat_xxz_raw_12.pdf}
\caption{}
\end{subfigure}
\hfill
\begin{subfigure}{0.47\textwidth}
\centering
\includegraphics[width=\linewidth]{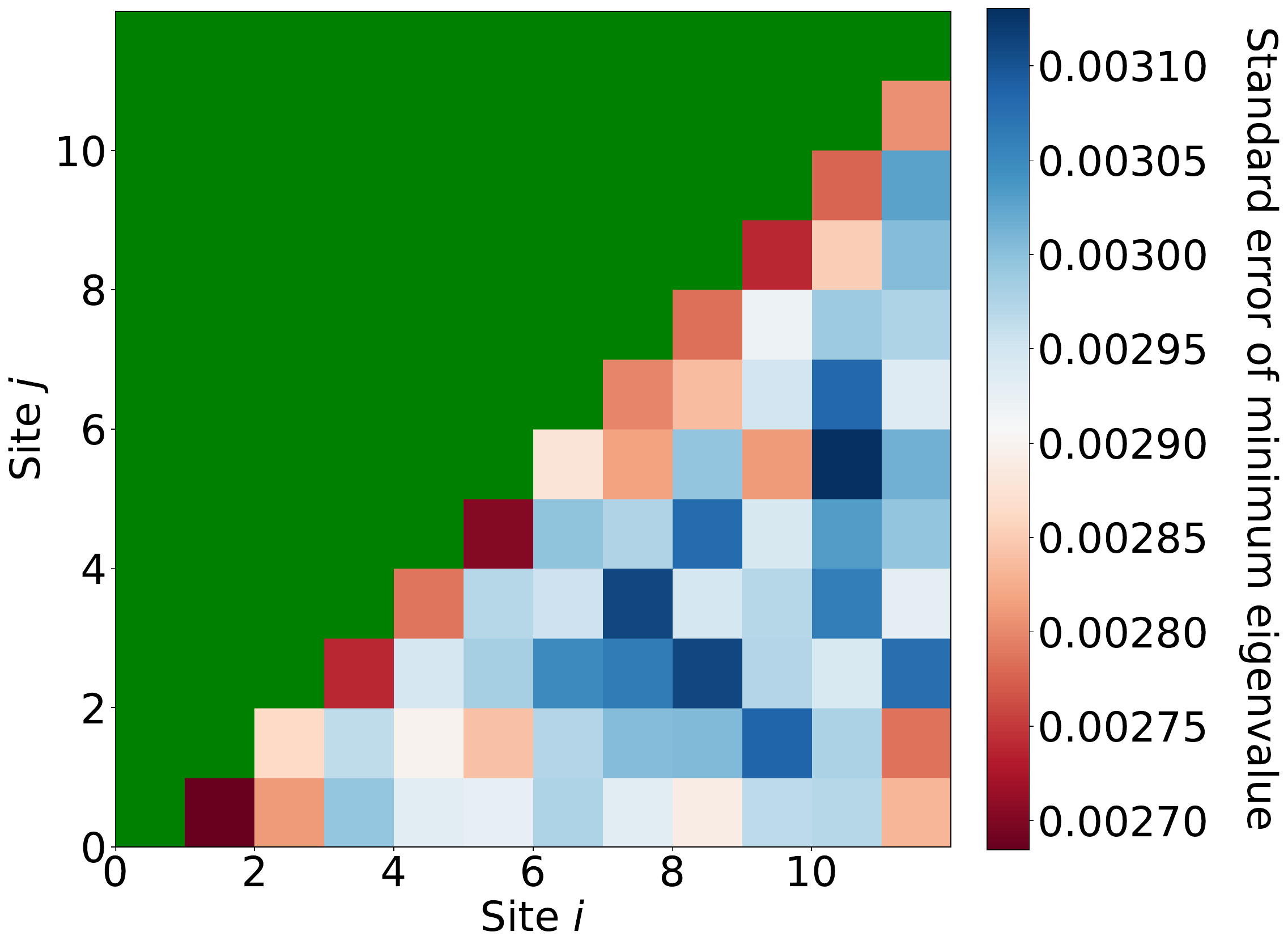}
\caption{}
\end{subfigure}
\vspace{0.3cm}
\begin{subfigure}{0.48\textwidth}
\centering
\includegraphics[width=\linewidth]{lmin_heat_xxz_miti_12.pdf}
\caption{}
\end{subfigure}
\hfill
\begin{subfigure}{0.47\textwidth}
\centering
\includegraphics[width=\linewidth]{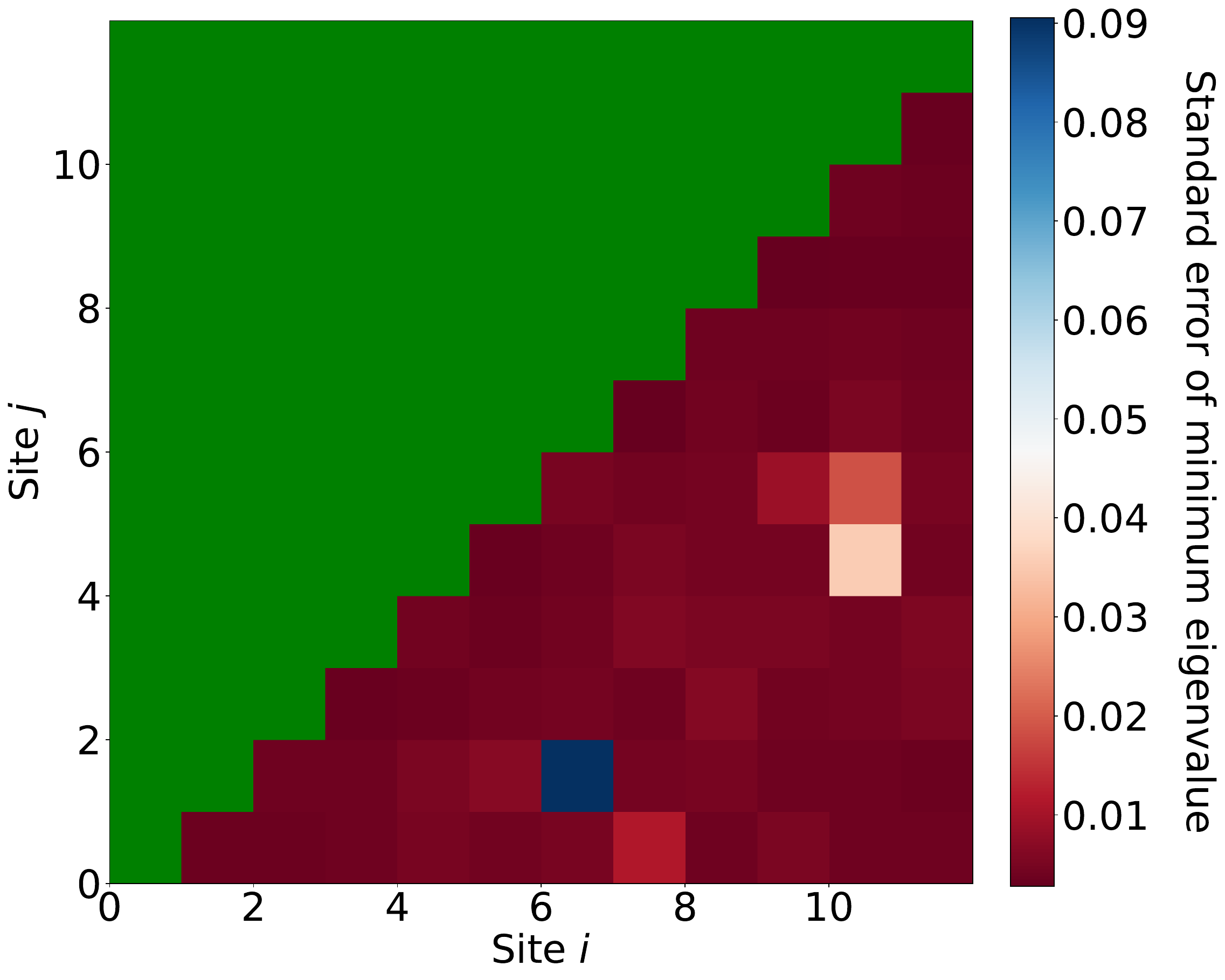}
\caption{}
\end{subfigure}
\caption{ Heatmaps (a,c,e) and corresponding error maps (b,d,f) for $N=12$ XXZ model. (a,b) Correlation function $C(X,Y)=(\langle X_i X_{i+d} \rangle + \langle Y_i Y_{i+d} \rangle)/2.0$. (c,d) Unmitigated $\lambda_\text{min}$. (e,f) $\lambda_\text{min}$ after error mitigation. }

\label{fig:xxz_12_heatmap_error}

\end{figure*}
\begin{figure*}[!htbp]
\centering
\begin{subfigure}{0.48\textwidth}
\centering
\includegraphics[width=\linewidth]{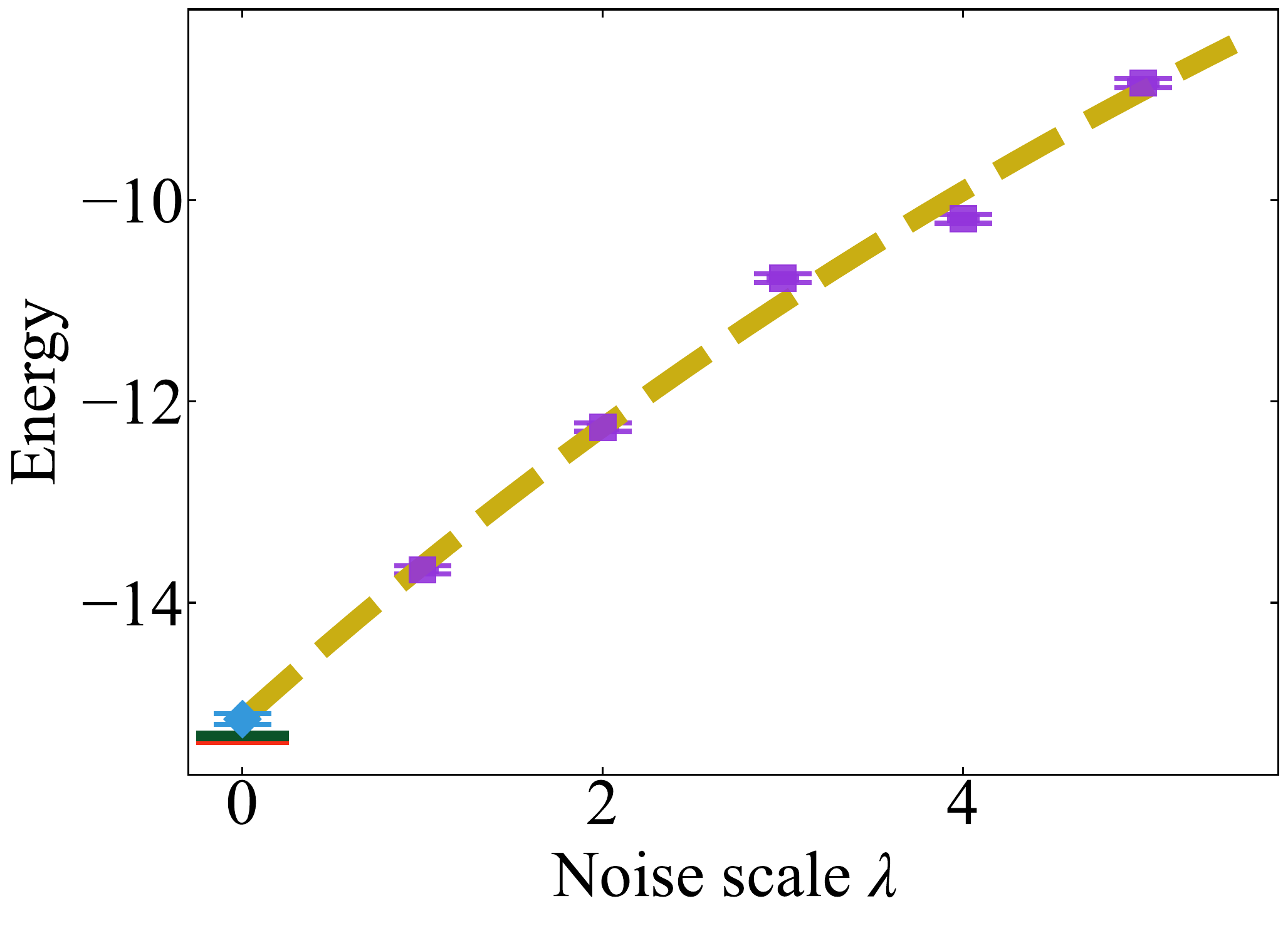}
\end{subfigure}
\hfill
\begin{subfigure}{0.48\textwidth}
\centering
\includegraphics[width=\linewidth]{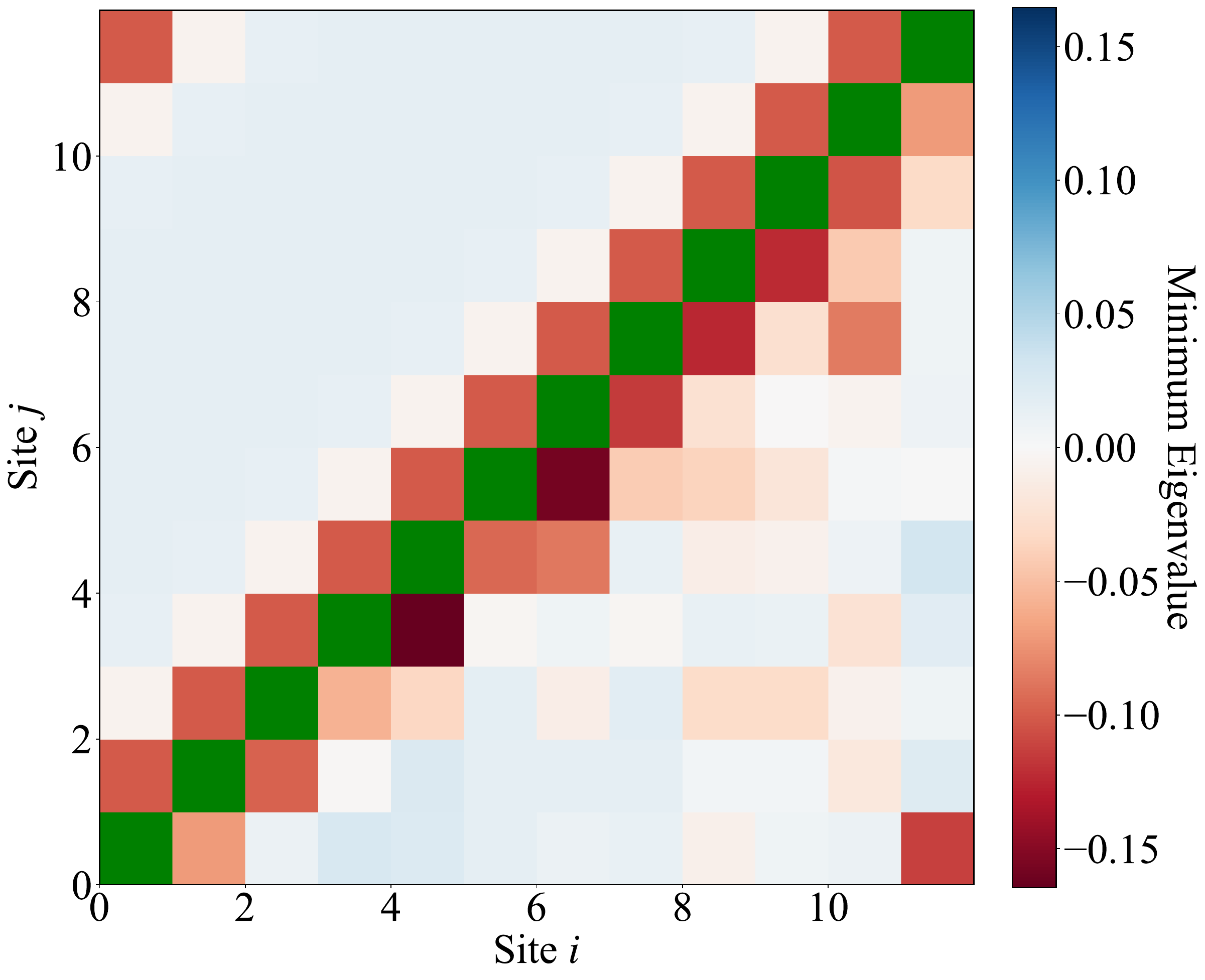}
\end{subfigure}
\vspace{0.3cm}
\begin{subfigure}{0.48\textwidth}
\centering
\includegraphics[width=\linewidth]{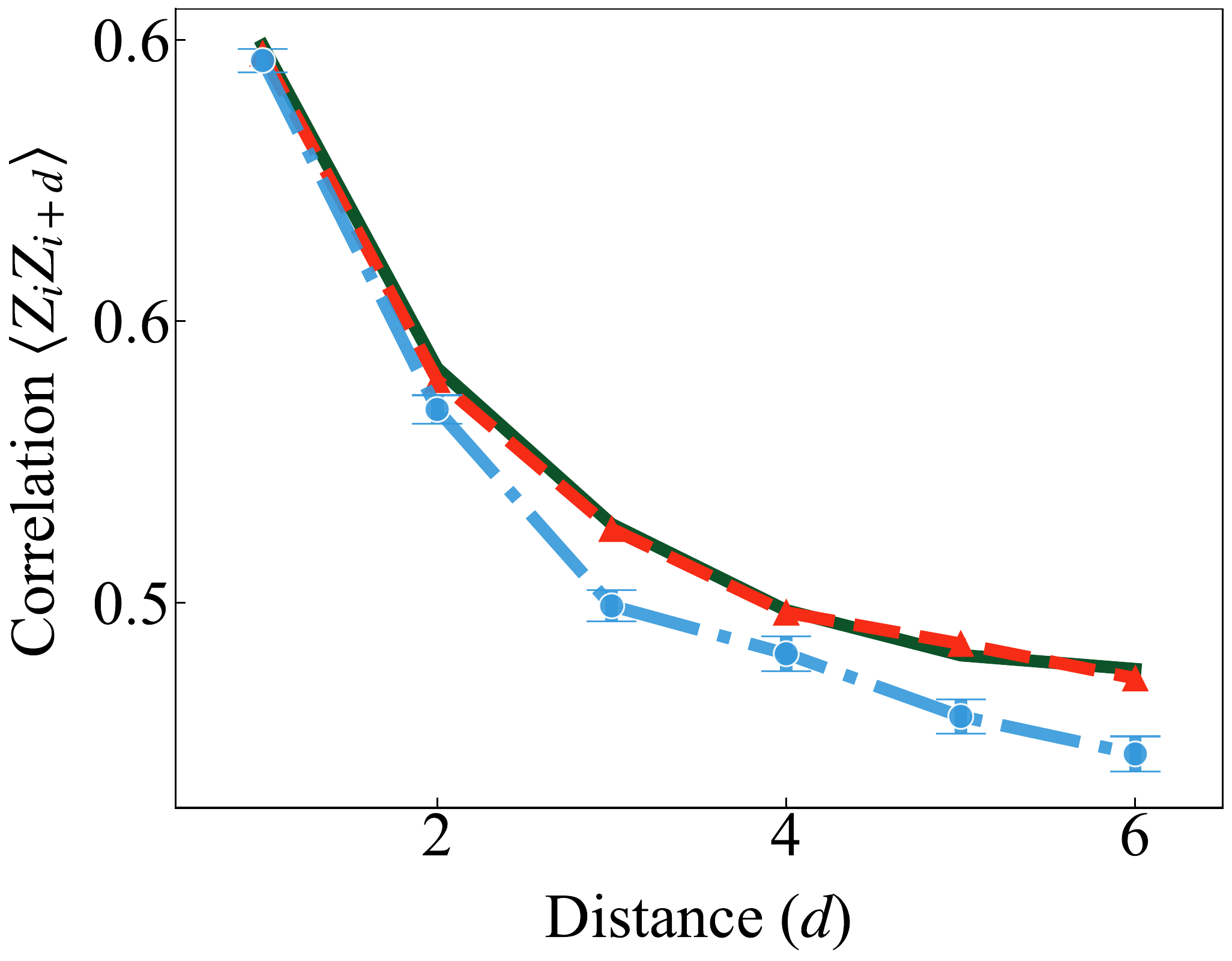}
\end{subfigure}
\hfill
\begin{subfigure}{0.48\textwidth}
\centering
\includegraphics[width=\linewidth]{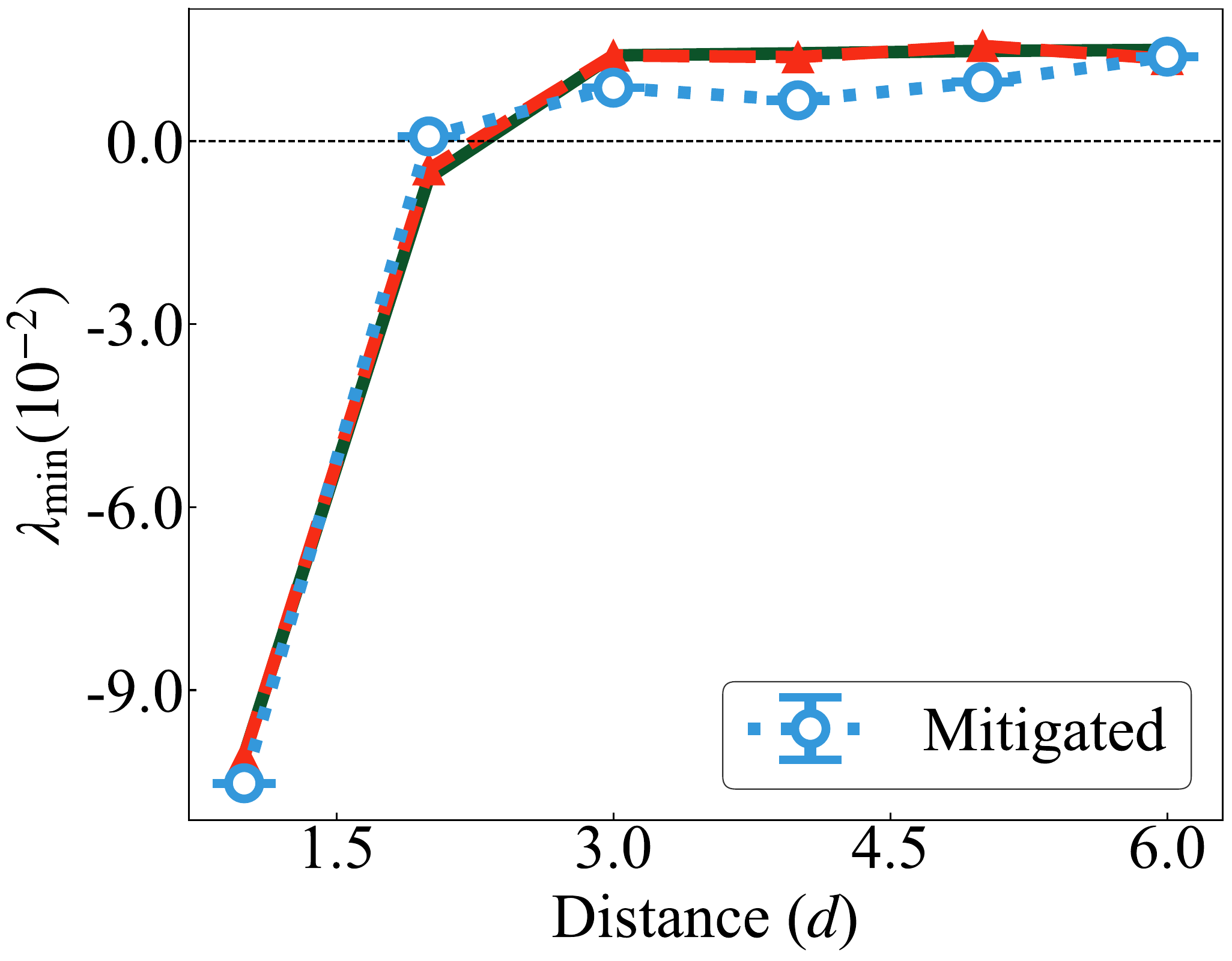}
\end{subfigure}
\caption{
Results for $N=12$ TFIM model. }

\label{fig:tfim_12}

\end{figure*}
\begin{figure*}[!htbp]
\centering
\begin{subfigure}{0.48\textwidth}
\centering
\includegraphics[width=\linewidth]{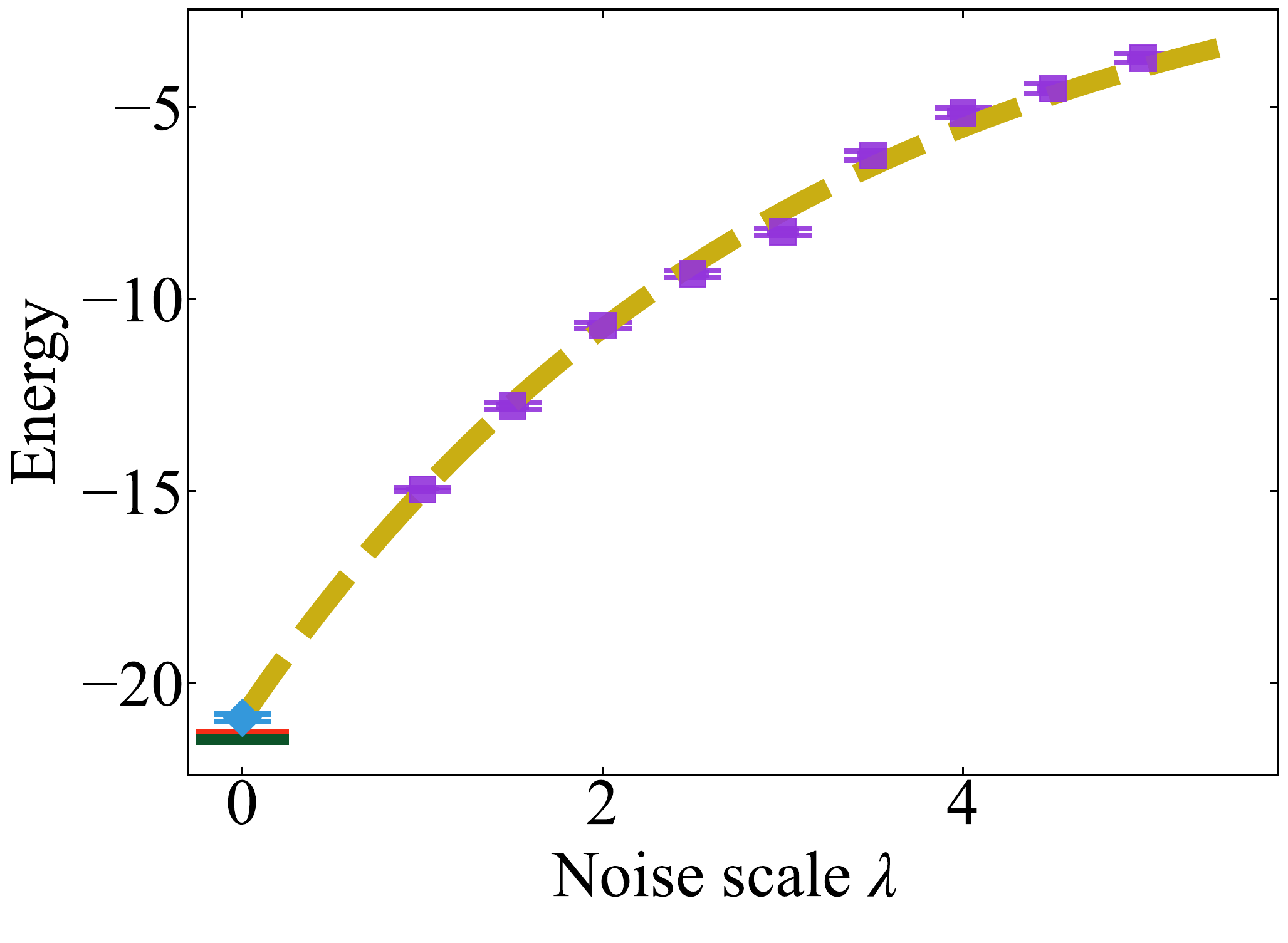}
\end{subfigure}
\hfill
\begin{subfigure}{0.48\textwidth}
\centering
\includegraphics[width=\linewidth]{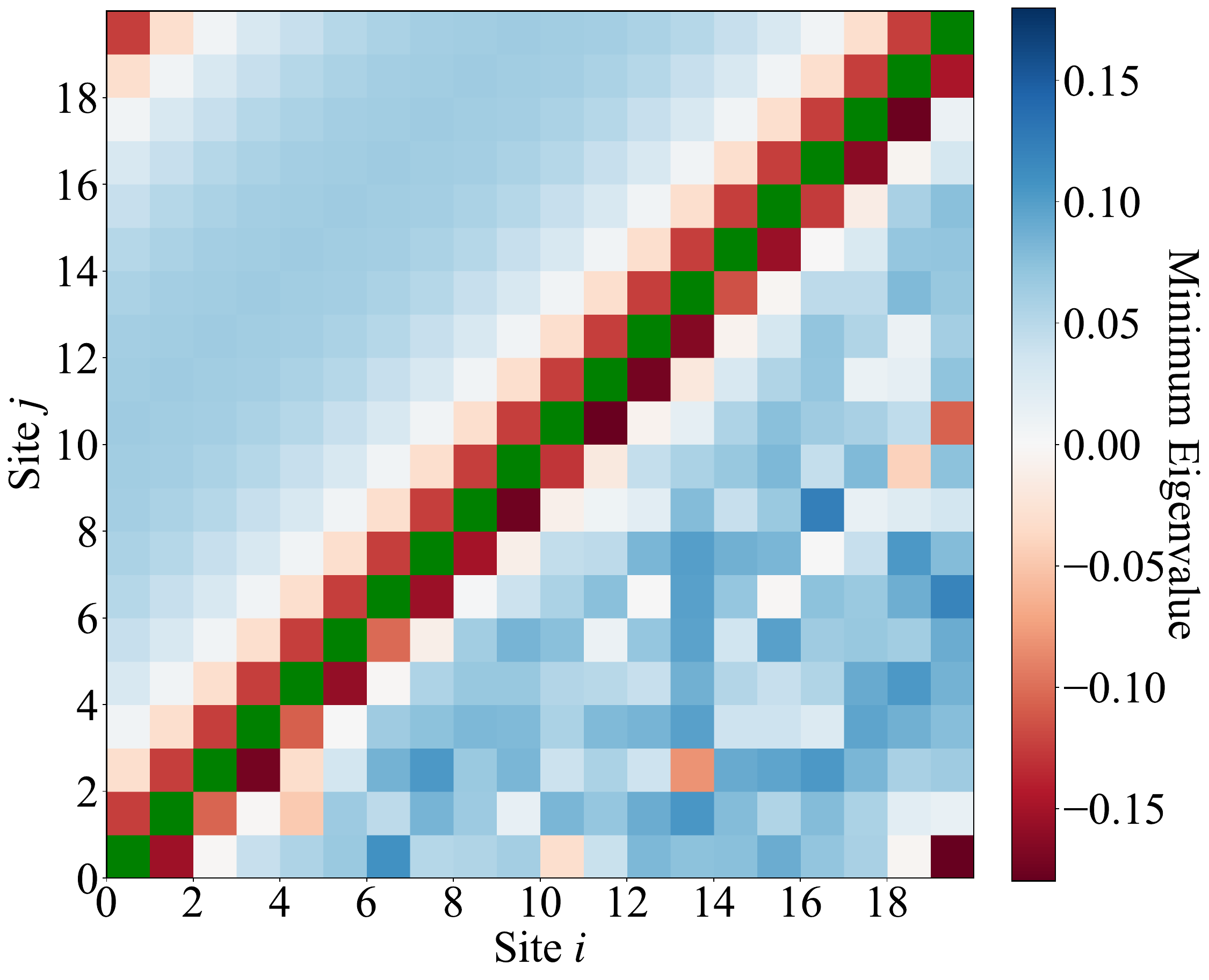}
\end{subfigure}
\vspace{0.3cm}
\begin{subfigure}{0.48\textwidth}
\centering
\includegraphics[width=\linewidth]{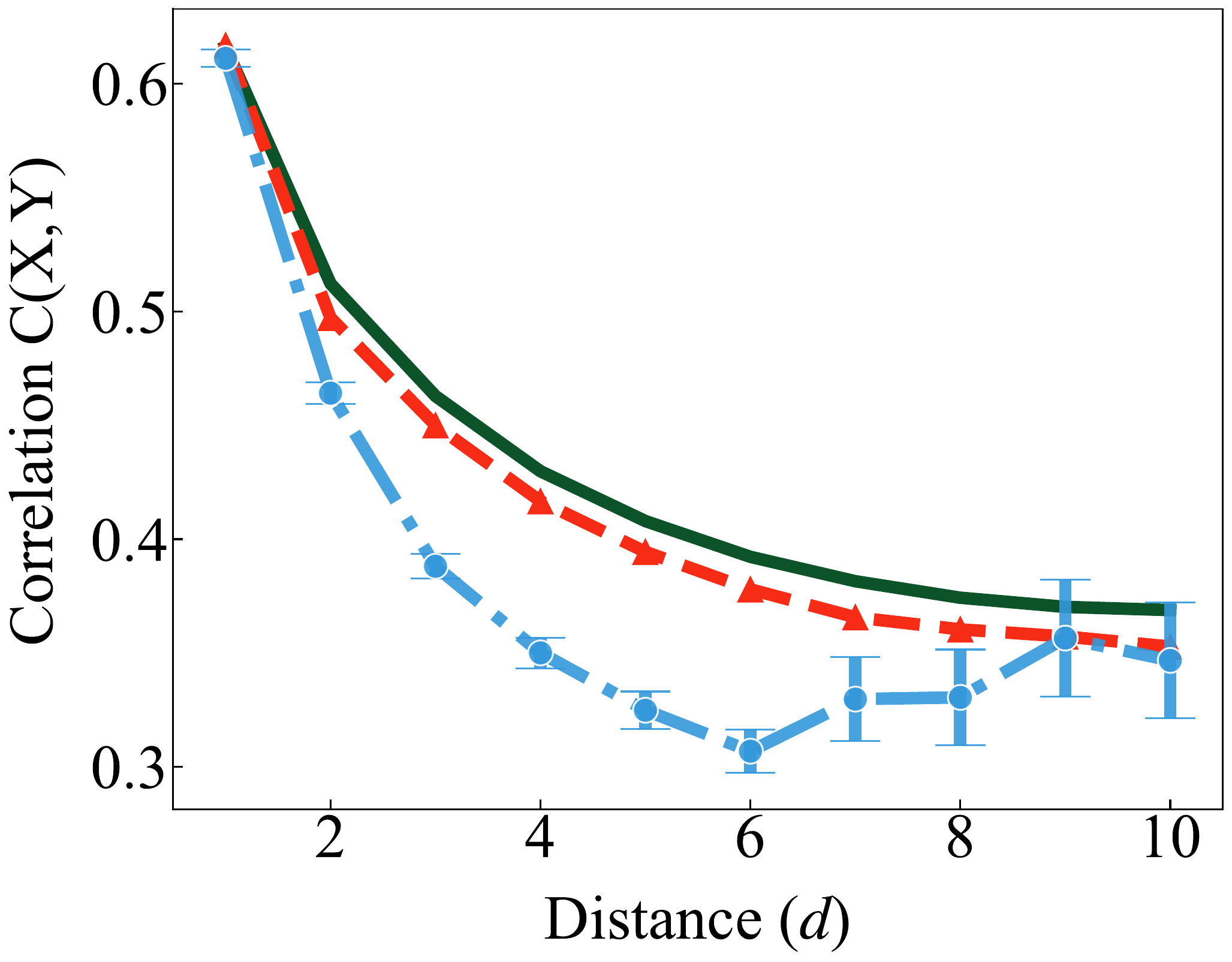}
\end{subfigure}
\hfill
\begin{subfigure}{0.48\textwidth}
\centering
\includegraphics[width=\linewidth]{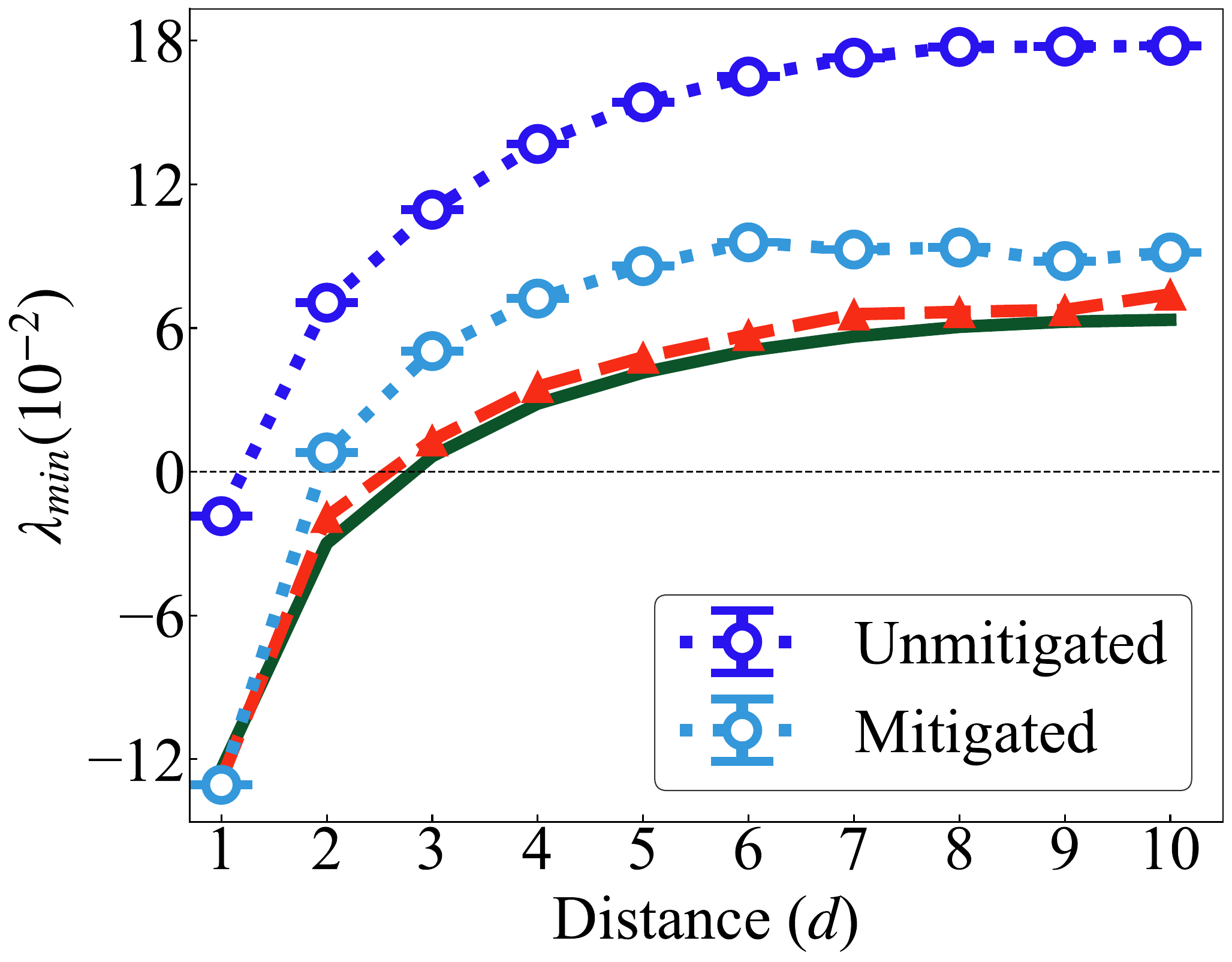}
\end{subfigure}
\caption{
Results for $N=20$ XXZ model. }

\label{fig:xxz_20}

\end{figure*}

\end{document}